\documentclass[apj]{emulateapj}

\usepackage{natbib}
\usepackage{color}
\usepackage{graphicx}
\bibpunct{(}{)}{;}{a}{}{,}

\begin{document}

   \title{
Local tadpole galaxies: dynamics and  metallicity
}
   \author{
          	J.~S\'anchez~Almeida\altaffilmark{1,2},
        	C.~Mu\~noz-Tu\~n\'on\altaffilmark{1,2}, 
                D.~M.~Elmegreen\altaffilmark{3}, 
                B.~G.~Elmegreen\altaffilmark{4},\\
                and
         	J.~M\'endez-Abreu\altaffilmark{1,2}
          }
\altaffiltext{1}{Instituto de Astrof\'\i sica de Canarias, E-38205 La Laguna,
Tenerife, Spain}
\altaffiltext{2}{Departamento de Astrof\'\i sica, Universidad de La Laguna,
Tenerife, Spain}
\altaffiltext{3}{Department of Physics and Astronomy, Vassar College, 
Poughkeepsie, NY 12604, USA}
\altaffiltext{4}{IBM Research Division, T.J. Watson Research Center, Yorktown Heights, 
NY 10598, USA}
\email{jos@iac.es, cmt@iac.es, elmegreen@vassar.edu, bge@us.ibm.com, jairo@iac.es}%, \dots}
\begin{abstract}
Tadpole galaxies, with a bright peripheral clump on a faint tail, are 
morphological types unusual in the nearby universe but very common
early on.
Low mass local tadpoles were identified and studied photometrically  in a previous work,
which we complete here  analyzing their chemical and dynamical properties.
We measure H$\alpha$ velocity curves of seven local tadpoles, representing
50\,\% of the  initial sample.
Five of them show evidence for rotation ($\sim 70$\,\%), and  a sixth target hints  at it.
Often the center of rotation is spatially offset with respect to the tadpole 
head (3 out of 5 cases).
The size and velocity dispersion of the heads are typical of  giant HII regions,
and three of them yield dynamical masses in fair agreement
with their stellar masses as inferred from photometry. 
%The linewidths tend to decrease at the head.
%Thus the bright heads seem to be self-gravitating systems,
%with their random motions reduced with respect to the 
%rest of the galaxy.
%
In four cases the velocity dispersion at the head 
is reduced with respect to its immediate surroundings.  
The  oxygen metallicity estimated from [NII]$\lambda$6583/H$\alpha$ often
shows significant spatial variations across the galaxies
($\sim$0.5~dex), being smallest at the head 
and  larger elsewhere. The resulting  chemical abundance 
gradients are opposite to the ones observed in local spirals,
but agrees with disk galaxies at high redshift. 
We interpret the metallicity  variation as a sign of external 
gas accretion (cold-flows) onto the head of the tadpole.
%We interpret the metallicity gradient as a sign of 
%star formation driven by external gas accretion (cold-flows).     
%
The galaxies are low metallicity outliers of the mass-metallicity 
relationship.
In particular, two of the tadpole heads are extremely metal poor, 
with a metallicity smaller than a tenth of the solar value.
These two targets are also very young (ages smaller than 5\,Myr). 
All these results combined are consistent with the local 
tadpole galaxies being  disks in early stages of assembling,
with their star formation sustained by accretion of external metal poor gas.
\end{abstract}
   \keywords{
     galaxies: abundances --
     galaxies: dwarf --
     galaxies: evolution --
     galaxies: formation --
     galaxies: kinematics and dynamics --
     galaxies: structure
               }
\slugcomment{Accepted version}%\@\dots/paper81/ver3/}

% \shorttitle{}
% \shortauthors{}

%%%%%%%%%%%%%%%%%%

\section{Introduction}\label{introduction}

Tadpole galaxies consist of a large star-forming clump at one
end (the {\em head}) and a long diffuse region to one side 
(the {\em tail}). This asymmetric 
morphology is rather common at high redshift but rare in 
the local universe. For example, tadpoles constitute 10\% of all 
galaxies larger than 10 pixels in the Hubble Ultra Deep Field (UDF) 
\citep{2007ApJ...658..763E,2010ApJ...722.1895E}, 
and they represent 6\% of the UDF galaxies
identified  by \citet{2006ApJ...639..724S} and \citet{2006NewAR..50..821W}
using automated search algorithms. 
In contrast, \citet[][hereafter Paper I]{2012ApJ...750...95E}
find only 0.2\%\ tadpoles among the star-forming
local galaxies of the Kiso survey by \citet{2010PNAOJ..13....9M}.
This decrease suggests the tadpole morphology to represent
a common but transit phase during the assembly of some
galaxies. Since local tadpole galaxies are very low mass objects 
compared to their high redshift analogues 
\citep{2012ApJ...750...95E},  the 
phase must be already over for the
local descendants of high redshift tadpoles.

Another independent observation also suggests that the tadpole 
morphology characterizes a very early phase of evolution. 
From the point of view of their chemical content, 
Extremely Metal Poor (XMP) galaxies  are the least evolved objects 
in the local universe 
\citep[e.g.,][]{1992MNRAS.255..325P,2000A&ARv..10....1K,2004ApJ...616..768I}. 
They represent only 0.1\,\% of the galaxies in an arbitrary
nearby volume \citep[e.g.,][]{2011ApJ...743...77M}, but
a significant fraction of these chemically primitive objects
turn out to have tadpole or cometary shape 
\citep[$\sim$75\,\%;][]{2008A&A...491..113P,2011ApJ...743...77M}.
This association between low metallicity and tadpole shape
suggests that they are attributes characteristic of very young systems.

The tadpole structure could have a variety of origins.
\citet{2010ApJ...722.1895E} showed lopsided ring-like 
galaxies that would look like tadpoles if viewed edge-on. 
Tadpoles could also result from mergers or close 
galaxy-galaxy interactions  
\citep[e.g.,][]{1982MNRAS.198..535B,
2005ApJ...629L..89C,
2006ApJ...639..724S,
2006NewAR..50..821W,
2010ApJ...725..353D}.
However, this merger scenario 
cannot be the universal pathway
to tadpole formation, 
at least not locally,  since the class of 
dwarf local tadpoles tends to be relatively isolated 
\citep{1993AJ....106.1784C,2000A&ARv..10....1K} 
and lacks obvious tidal features  
\citep[][]{2008A&A...491..113P}.
Another possibility is that the lopsided starburst results from
ram compression by motion through the intergalactic medium
\citep{2010ApJ...722.1895E}.
Tadpoles could be disks with star formation triggered 
on the leading side and the rest visible as a red tail 
of older stars. Alternatively, they could be heavily stripped galaxies 
with star formation and old stars at the leading edge, 
and a tail made from star formation in the stripped gas
\citep[e.g.,][]{2009AJ....138.1741C}. 
The scenario in which dwarf galaxies are converted into
dwarf spheroidal galaxies in the halos of larger galaxies
\citep{1983ApJ...266L..21L,1994ApJ...428..617V,
%2003AJ....125.1926G,
2006MNRAS.369.1021M} %,2011ApJ...739....5W},
or in galaxy clusters 
\citep{2008ApJ...674..742B,2004AJ....128..121V}
could involve a tadpole phase as the gas and young
stars are pulled behind.
A third possibility is that some tadpoles are normal disk galaxies
with a large turbulent Jeans length for gravitational collapse of
their interstellar medium. This happens in galaxies that have
either  small rotational velocities or large turbulent motions, and 
in this case only large star forming clumps can be produced 
\citep{2006ApJ...645.1062F,2009ApJ...701..306E}.
If there is only one clump at any particular time, then it
will appear as a tadpole viewed from the right perspective.
A fourth possibility is that the head-tail structure results from
propagating star formation across a disk 
\citep[e.g.,][]{1997ApJ...486L..75F,2008A&A...491..113P}.
Finally, as we propose in this paper,
the head of the tadpole may result from accretion of 
external flows of pristine gas, that penetrate the dark matter halo 
and hit and heat a pre-existing disk.
Cosmological simulations predict cold-flow buildup to be the 
main mode of galaxy formation 
\citep{2009Natur.457..451D,2012ApJ...745...11G}, 
and the incoming gas is expected to form giant clumps that
spiral in and merge into a central spheroid 
\citep{1999ApJ...514...77N,2008ApJ...687...59G,2008ApJ...688...67E}. 
In addition, inflow of low-metallicity gas seems to be suggested by  
distorted HI velocities and morphologies of some local star-forming
galaxies  \citep[see][and references therein]{2012MNRAS.419.1051L}.

It is important to realize that the different formation mechanisms mentioned
above are not mutually exclusive, but exhibit a large
degree of overlap  between them. For example, the
cold-flow accretion may be considered as a (minor-)merger accretion, 
and it may also be regarded as the interaction of the galaxy
with the (filamentary) intergalactic medium.

In this context of galaxy formation, we examined  a 
sample of fourteen tadpole galaxies in the local 
universe for comparison with high-redshift tadpoles (Paper~I). 
These local tadpoles seem to form a 
continuous sequence with the UDF tadpoles studied by 
\citet{2010ApJ...722.1895E}.  
With regards to their
photometric properties, local tadpoles occupy the
low mass end in sequences such as star formation,
surface density and mass-to-light ratio.
In addition, the radial intensity profiles of the local
tadpoles show an 
exponential decrease at large galactocentric distances, which
was interpreted as evidence for an underlying disk.    
The work in Paper~I is
%by \citet[][Paper~I]{2012ApJ...750...95E} is 
followed up  in the present paper.
In order to determine the dynamical properties and
metallicities of local tadpoles, we measure H$\alpha$ spectra along 
the head-tail direction in a
representative fraction of the original  sample. 
The results reported here show  the galaxies to be  
rotating structures with an unexpected metallicity pattern,
which may shed light into the nature of the tadpole morphology.

%%%%%%%%%
The paper is organized as follows:  Sect.~\ref{observations}
describes the observations, the main steps of the reduction,  
and the main properties of the resulting spectra.  
Section~\ref{determination} details the approximations 
used along the work to derive physical parameters from spectra. 
Section~\ref{dynamical_properties} analyzes the dynamical properties 
of the galaxies, their rotation curves (RCs) and linewidths.
These are used to infer dynamical masses, which 
are then compared with the stellar masses derived in Paper~I
(Sect.~\ref{dynmasses}).
Section~\ref{fluxes_and_else} studies the light profiles both 
in continuum and  H$\alpha$ emission.
The variation of the oxygen abundance along the 
galaxy is measured in Sect.~\ref{metalcont}.
Notes on %specificities of 
individual galaxies are given in 
Sect.~\ref{notes_on_galax}.
The results of our observation in the context of tadpoles as 
dynamically young disks  are discussed in Sect.~\ref{discussion}.
%

%%%%%%%%%%%%%%%%%%%%
%
\section{Observations and data reduction}\label{observations}
\begin{figure*}
\includegraphics[width=0.3\textwidth]{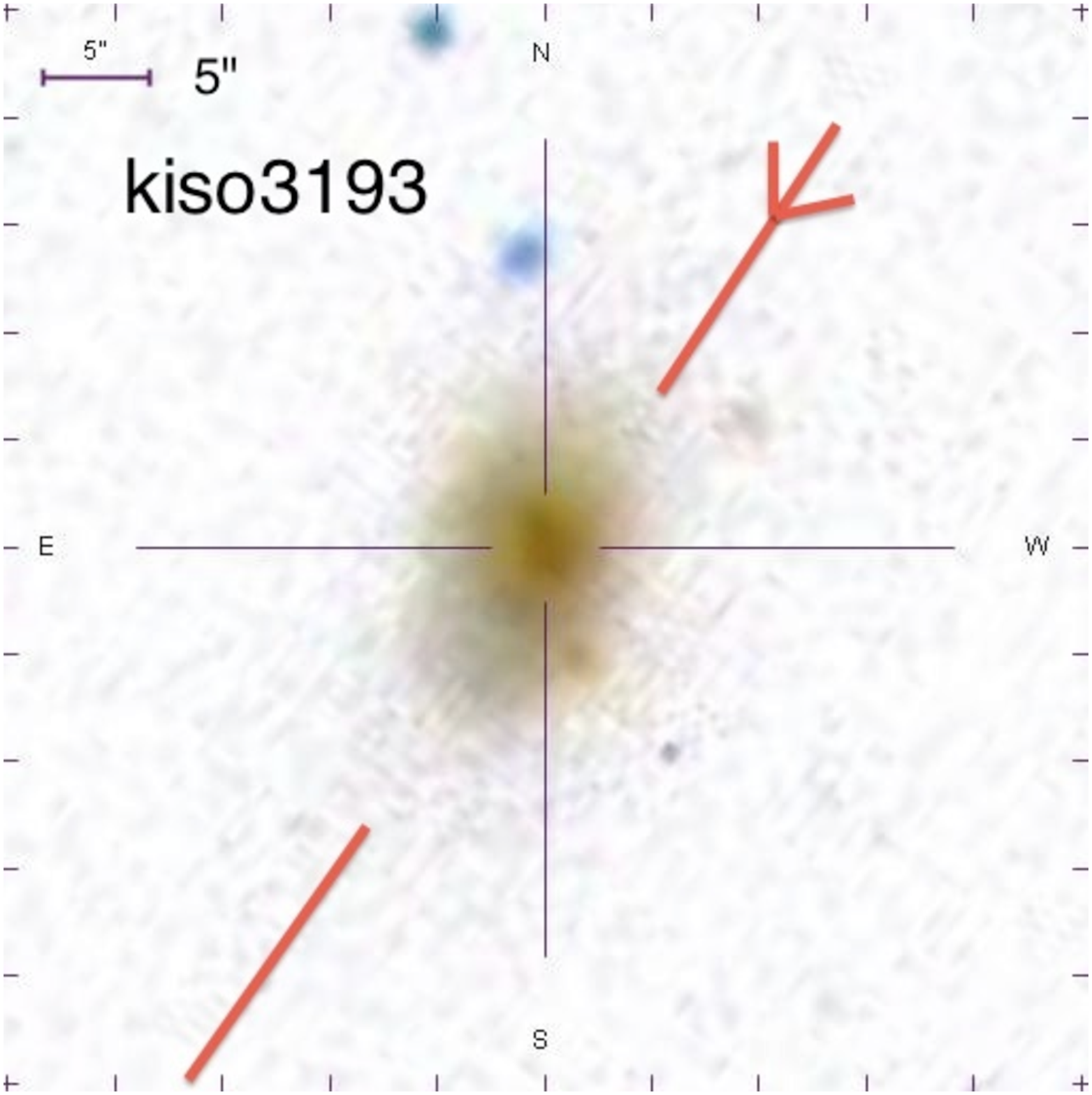}
\includegraphics[width=0.3\textwidth]{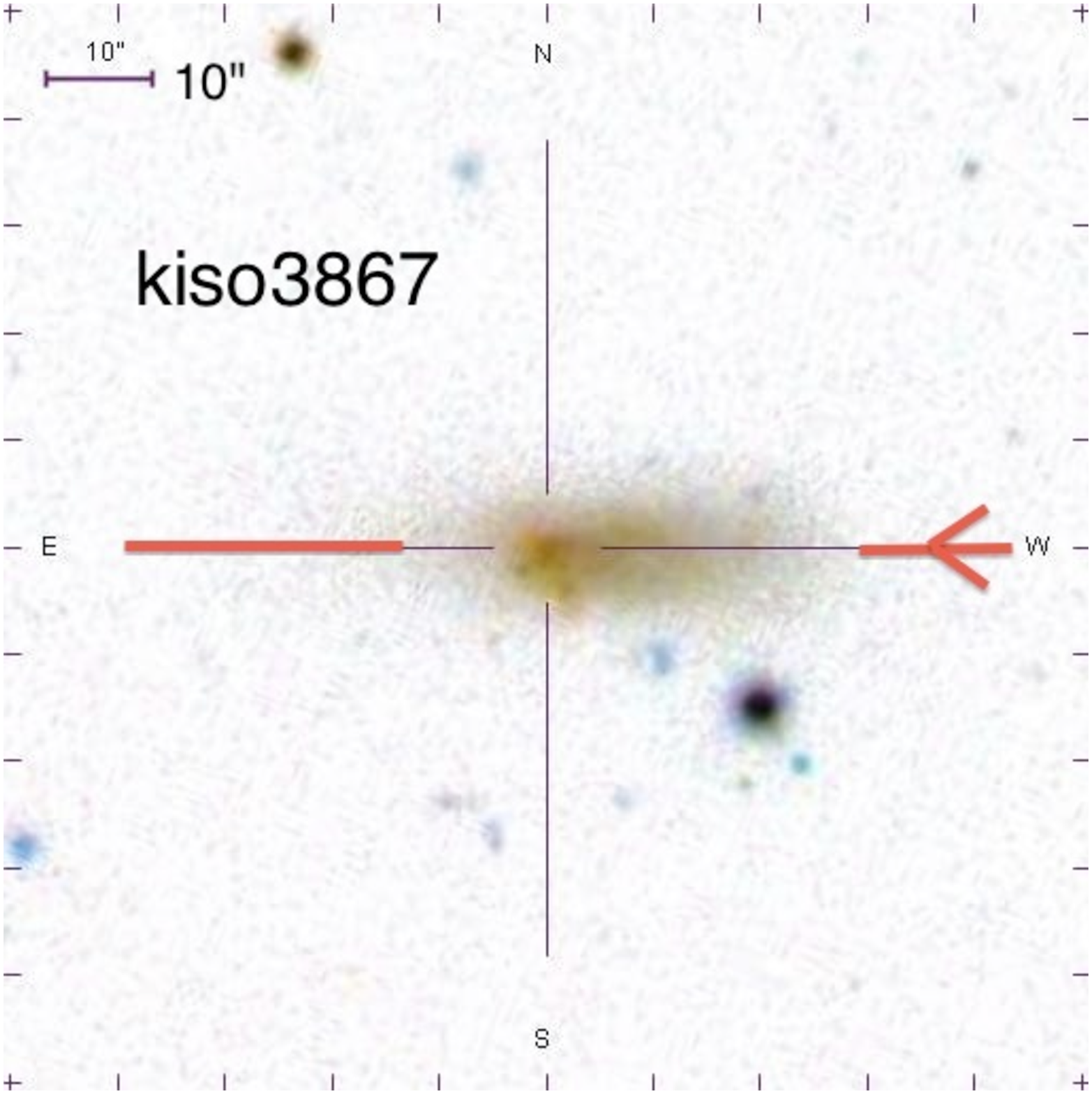}
\includegraphics[width=0.3\textwidth]{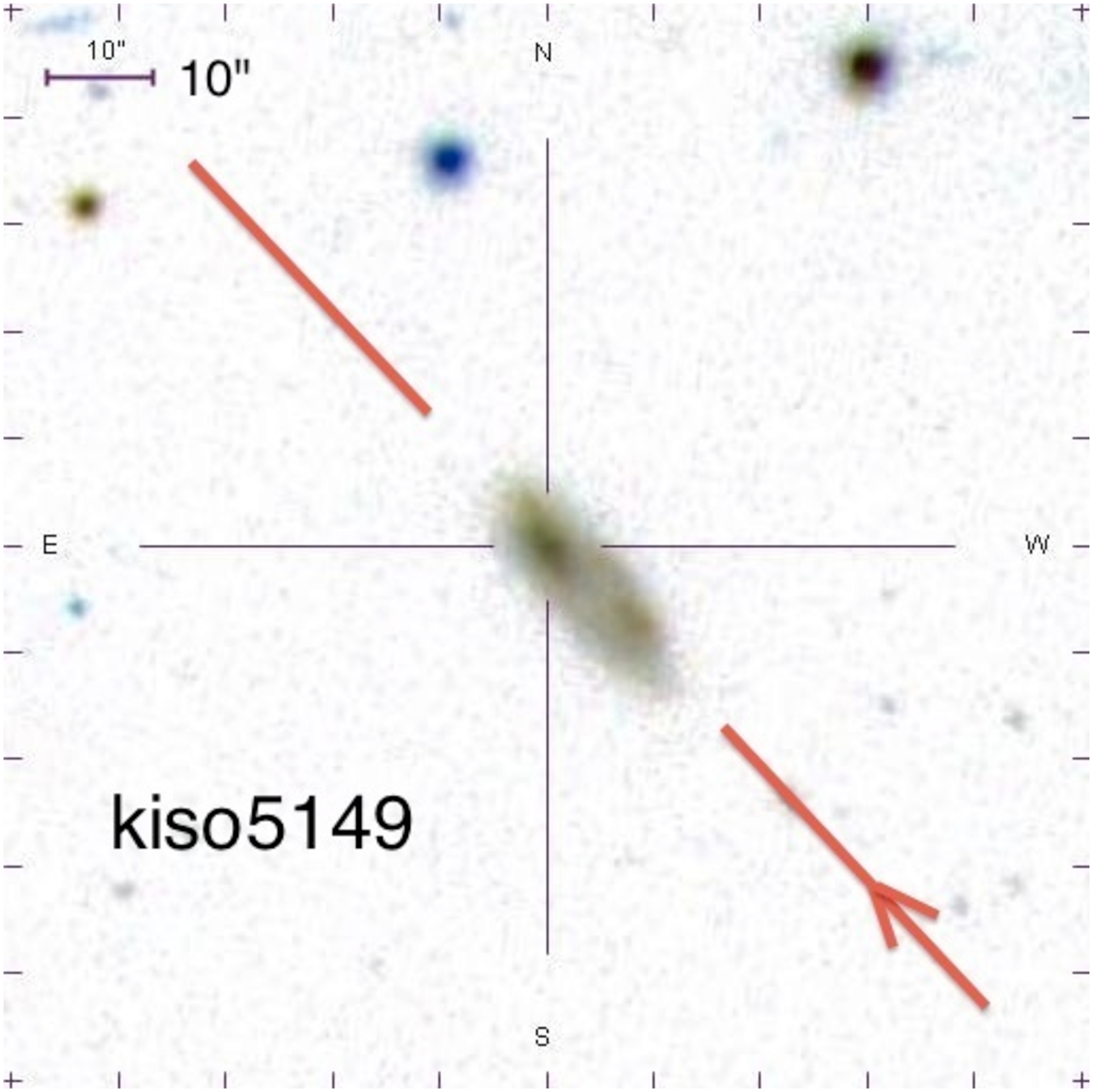}\hfill\break
\includegraphics[width=0.3\textwidth]{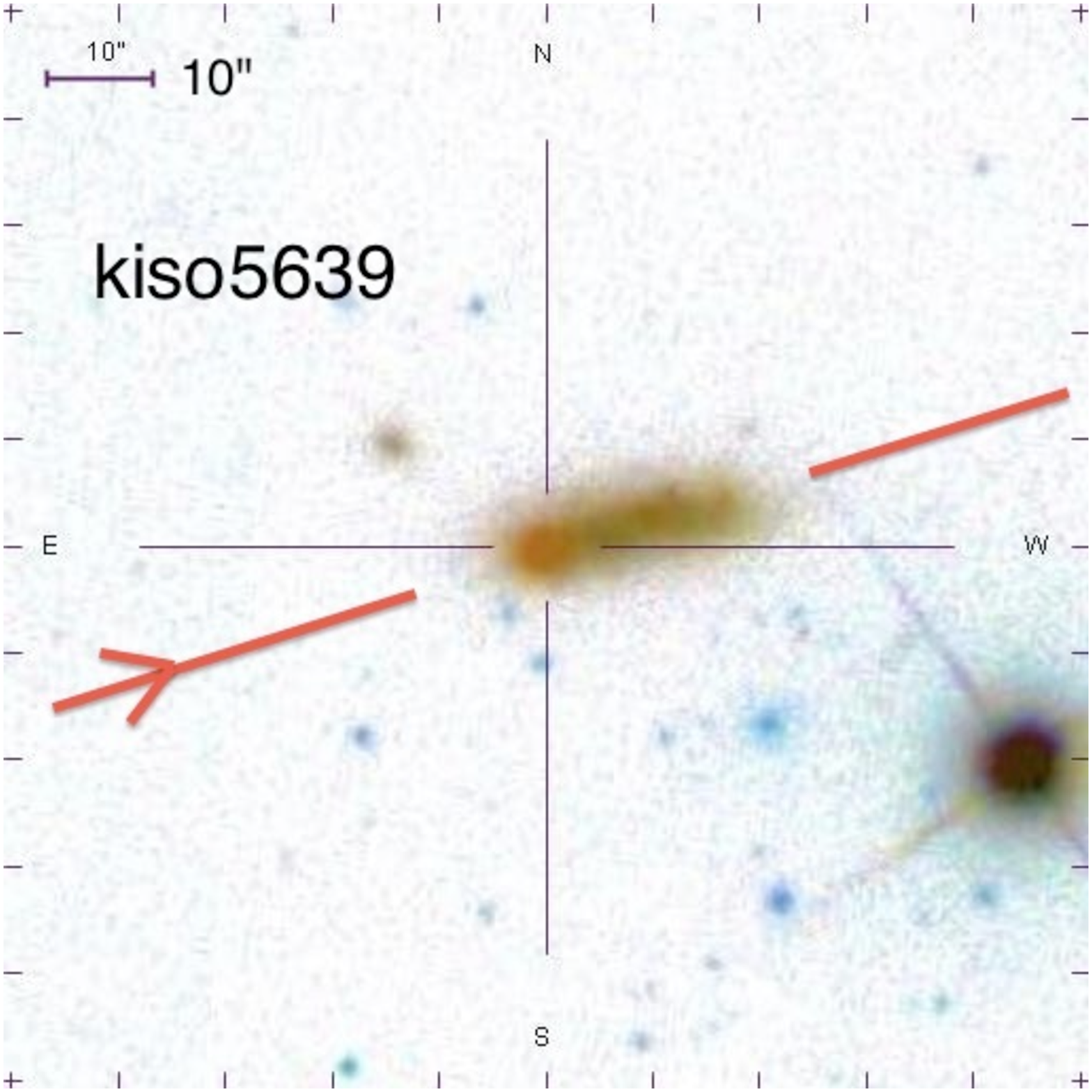}
\includegraphics[width=0.3\textwidth]{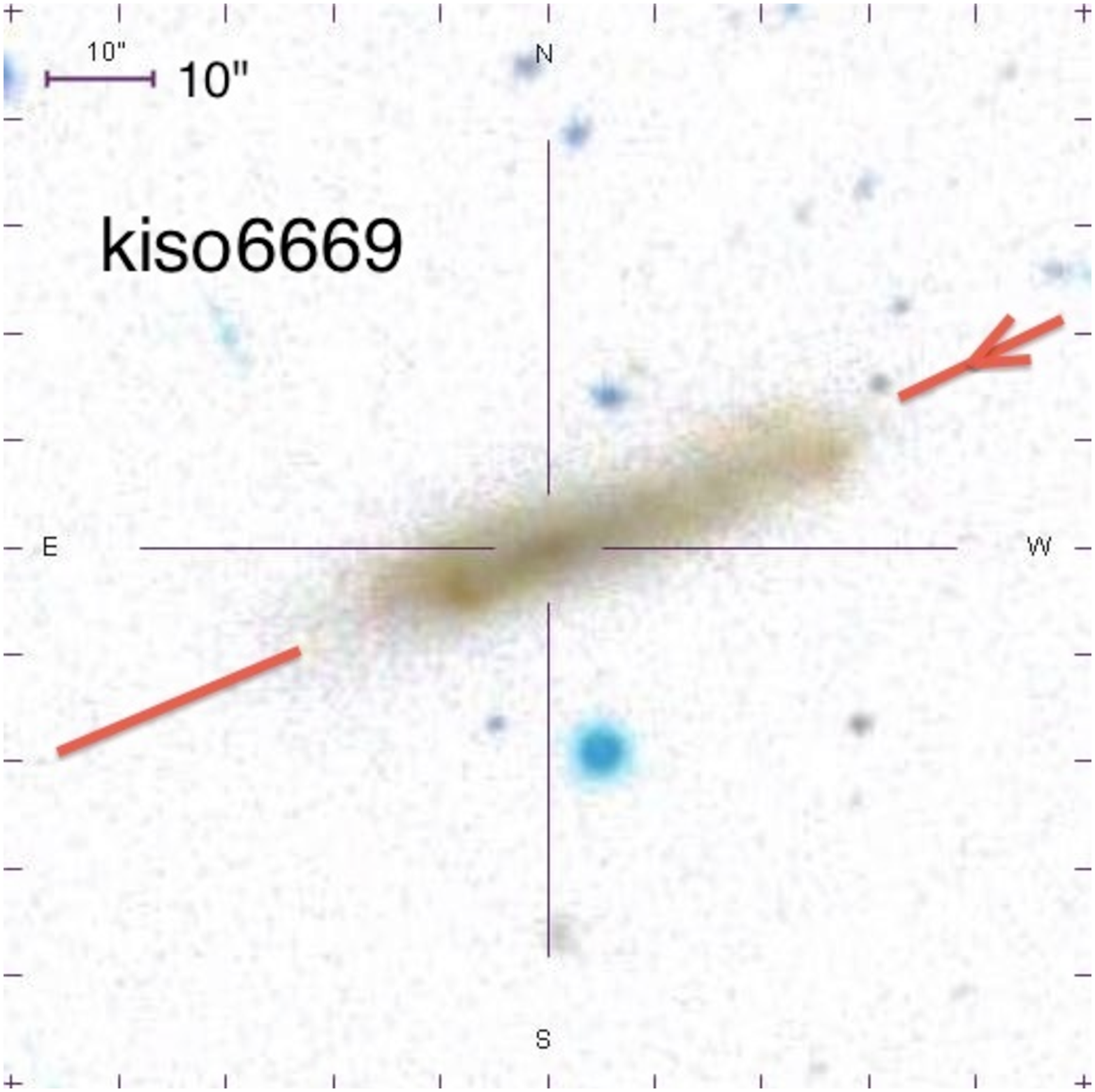}
\includegraphics[width=0.3\textwidth]{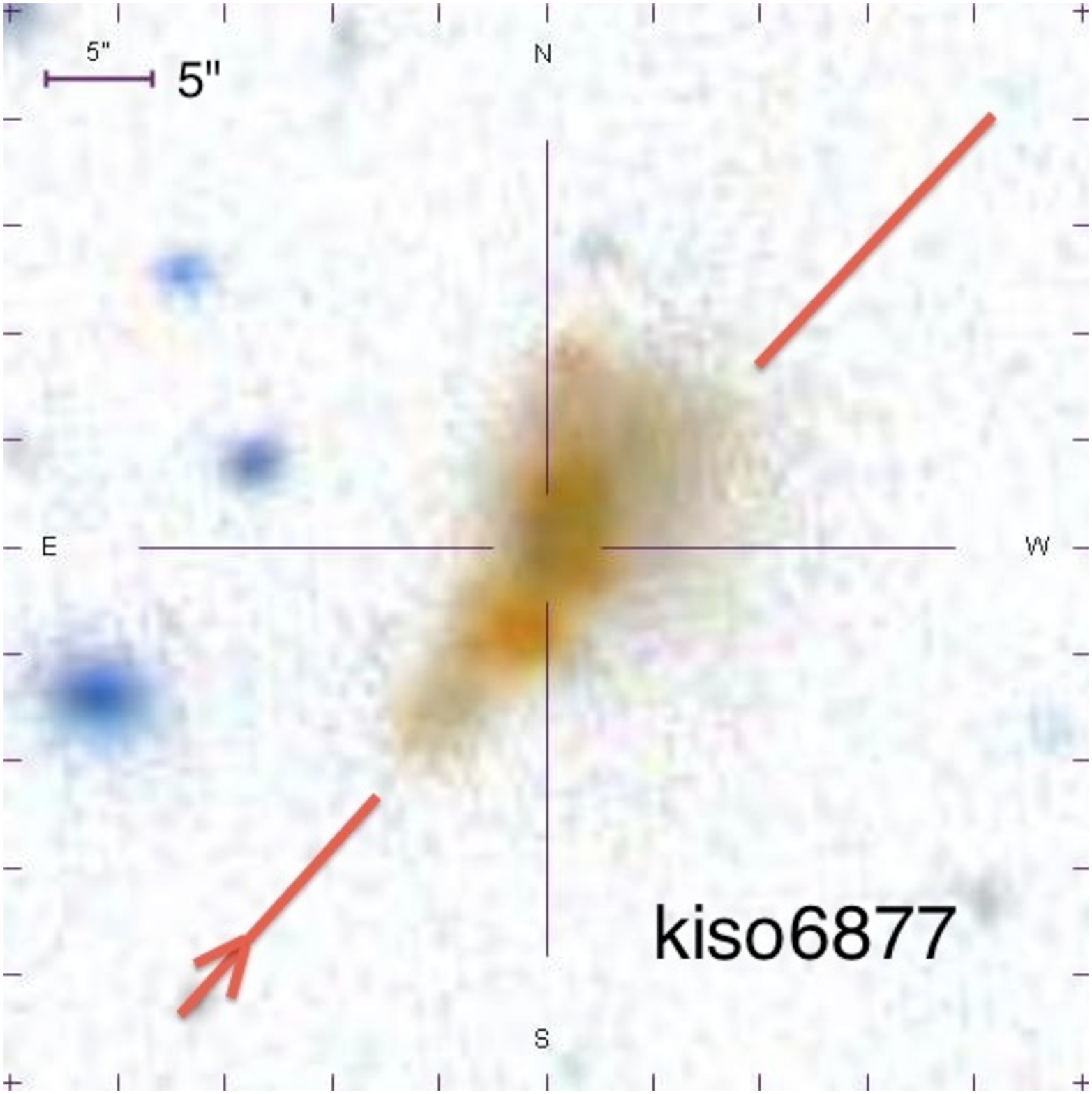}\hfill\break
\includegraphics[width=0.3\textwidth]{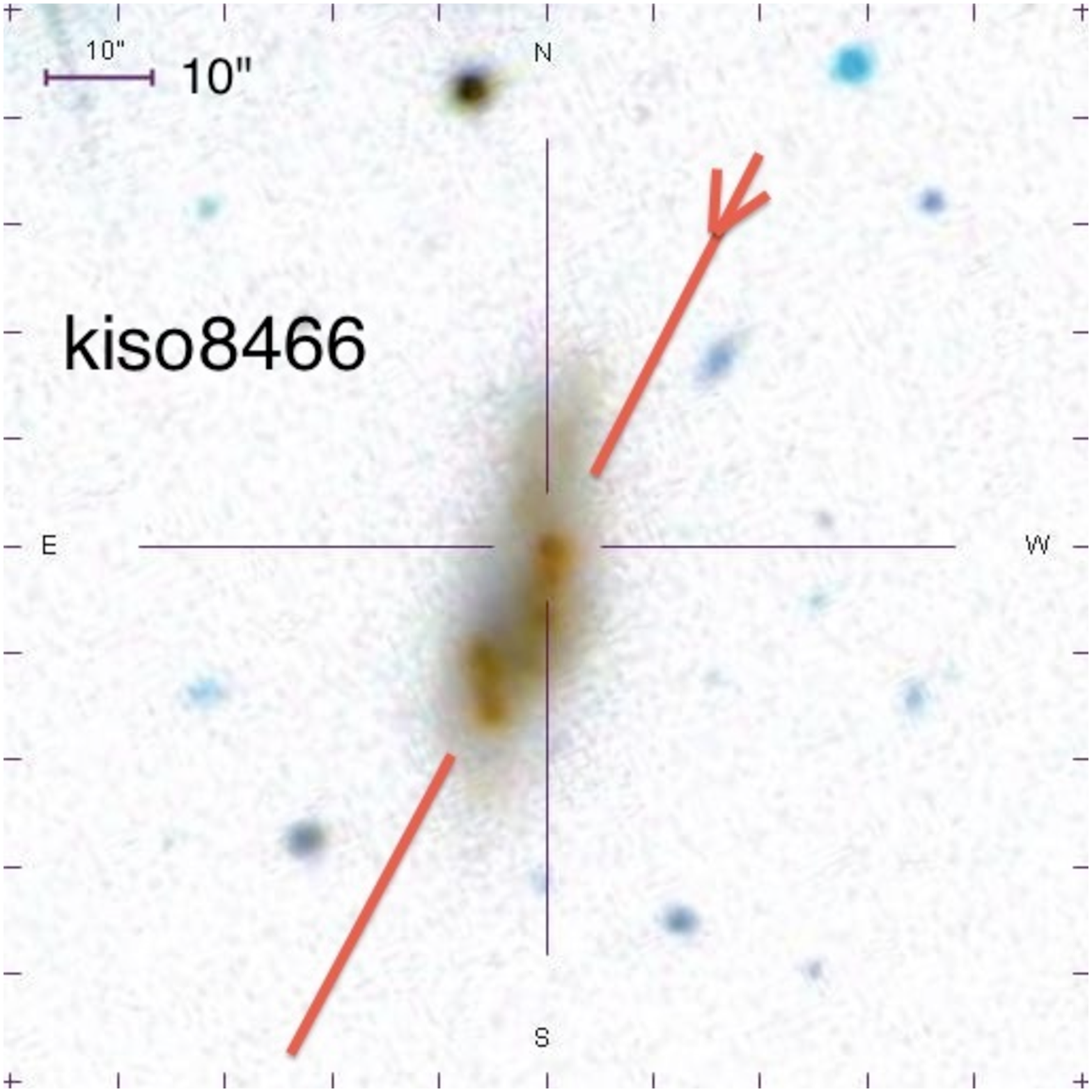}
\caption{
Images of the observed galaxies with the red line 
showing the orientation of the slit
and with the arrow indicating the sense of increasing distance 
along the slit used in the paper. 
The images have been taken from the SDSS database, and are
displayed with an inverted color palette so that the background sky looks white,
and the intrinsically blue galaxies appear reddish.  
The horizontal scales on the upper left
corner of the panels correspond to 5\arcsec or 10\arcsec as indicated.
}
\label{images}
\end{figure*}

The sample of local tadpole galaxies selected in Paper~I 
comes from the Kiso survey of UV intense galaxies 
\citep{2010PNAOJ..13....9M}.
Among those galaxies labelled in the Kiso catalog as clumpy
or having conspicuous HII regions, we visually inspected 
158 with images in SDSS-DR7 \citep{2009ApJS..182..543A}. 
Then tadpoles were subjectively selected as those 
objects with lopsided light distribution, so that the brightest 
clump is far to one end  and the rest of the galaxy is mostly 
featureless. This inspection rendered only 13 targets, 
which we completed with
one additional galaxy with similar features
from the University of Michigan survey of emission-line 
objects \citep{1978ApJS...36..587M}. Half of the sample
in Paper~I  was randomly selected for the follow-up 
work presented in  this paper.
The observing program was aimed at determining the dynamical properties 
of the galaxies based on long slit spectroscopic 
observations along the well-defined head-tail direction (see Fig.~\ref{images}).  
The project was planned for  
the spectrograph IDS  of 
the 2.5\,m Isaac Newton Telescope (INT)  at the Observatorio del Roque
de Los Muchachos 
\citep[][]{1985VA.....28..483L}.
%\citep[Jones et a. Vistas in Astronomy, vol28, 2, 1985][]{??}.  
It provides adequate
spectral resolution to determine centroids of emission lines with 
an accuracy better than 10\,km\,s$^{-1}$, 
which suffices to characterize RCs
even for dwarf galaxies. The original 
project could not be completed due to poor
 weather conditions, so it was finished using service time of
the 2.5\,m Nordic Optical Telescope (NOT), operated in the same 
observatory \citep[][]{1985ospn.book.....A}.

\begin{figure*}
\includegraphics[width=0.7\textwidth,angle=90]{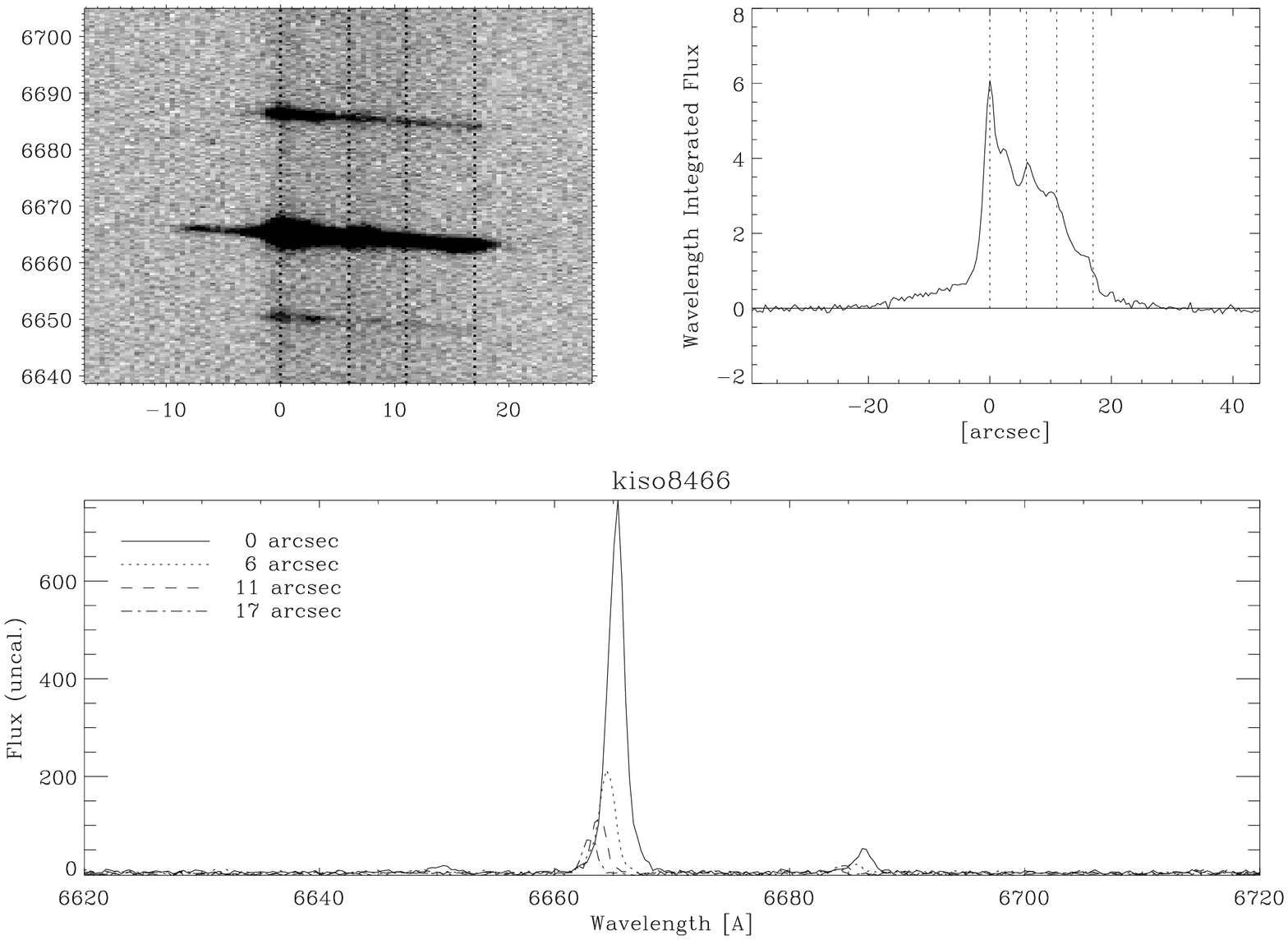}
\caption{
Example of a reduced spectrum from IDS@INT.
The image shows a negative of the spectrum
after reduction -- x-axis  represents position and y-axis wavelength.
Positions are given in arcsec from the galaxy head, whereas 
wavelengths are in \AA .
The  wavelength integrated flux (spectral flux) is shown in the
top right corner. 
Spectra corresponding to various positions
are shown in the bottom panel,
with the positions  indicated
as vertical dotted lines in the two upper plots.
The example corresponds to {\sc kiso5149}, but it is representative
of the full set.
The largest line is H$\alpha$ whereas the second largest is 
[NII]$\lambda$6583, used in our metallicity estimates.
}
\label{reduction}
\end{figure*}

The observing logbook is summarized in  Table~\ref{logbook}. 
The used IDS@INT setup includes a grating with 1800~lines\,mm$^{-1}$, 
which provides a scale in the spectrograph focal plane of  
0.32\,\AA\,pix$^{-1}$ equivalent to  0\farcs44\,pix$^{-1}$. The 
spectral resolution is set by the width of the 1\arcsec\ slit, 
which corresponds to 2.2\,pix, or 0.73\,\AA ,
or 33 km\,s$^{-1}$ in H$\alpha$.  
The camera covers some 700\,\AA\ around the central wavelength,
which was tuned to H$\alpha$, and which automatically included
the line [NII]$\lambda$6583 used in our metallicity estimates (Sect.~\ref{determination}).
The tadpoles {\sc kiso3193}, {\sc kiso3867}, {\sc kiso5149} and {\sc kiso8466}
were observed with this setup (Table~\ref{logbook}).
Figure~\ref{reduction} shows one of them as an example.  The
image contains the spectral region around H$\alpha$.
The reduction procedure included the standard bias and flatfield 
corrections, cosmic ray elimination, 
as well as removal of sky emission lines. 
%No traces of sky lines remain. 
The typical signal-to-noise (S/N) ratio in the peak H$\alpha$ emission 
exceeds 500,  and it decreases down to one in the outskirts of the
galaxy (Fig.~\ref{reduction}, bottom panel). This S/N 
is achieved after integrating for some one and a half hours 
(Table~\ref{logbook}).
Seeing was of the order of 1\arcsec\  (Table~\ref{logbook}).

The other three galaxies, {\sc kiso5639}, {\sc kiso6669}, and {\sc kiso6877}
were observed with the NOT telescope with
an observational setup equivalent to the previous one.
We used the  ALFOSC spectrograph % operated at the NOT telescope 
which, together with a high resolution grism, 
provided a scale of  0.26\,\AA\,pix$^{-1} \times$ 0\farcs 19\,pix$^{-1}$
at H$\alpha$. This pixel size over-samples the spectrum, whose resolution
was set by the 0\farcs 9 wide slit. In order
to match the pixel size to the actual resolution, the original data were  
re-binned $2\times 2$,  yielding the final resolution indicated in 
Table~\ref{logbook}.  The data reduction followed the standard process 
mentioned above, with a small difference worthwhile mentioning.
The standard procedure includes using images of the sky taken during 
twilight to characterize (and then to correct for) large-scale illumination
gradients in the images.  Such sky-flat images corresponding 
to  {\sc kiso6877} were not available. For reduction we used those from 
the previous night,  but this rendered spectra with small residual
sky lines. 
%In order to eliminate this problem as a source of uncertainty, 
%the reduction was independently repeated by two of us (JSA and JMA).
To estimate the uncertainties introduced by  this problem, 
the reduction  was independently repeated
by two of us (JSA and JMA). Thus different spectral regions
and  interpolations were  used to  construct  the flatfield  and to
subtract  the  sky  background.  The  two  reductions  are  in  good
agreement and provide consistent results.
%,  and considering  the  signal-to-noise of  the spectra,  the
%sky-flat problem does not affect the results in the work. 
Table~\ref{obs_summary} contains the final physical parameters 
inferred from the two parallel reductions,  and they 
agree to the point that their small differences do not affect
%The  two reductions 
the results in the work.  Seeing during observations was  
sub-arcsec. The NOT spectra have S/N similar to the INT spectra;
the spectral resolution is slightly  worse,
and the angular 
resolution slightly better.

Determining the actual spectral resolution is central for
the proper interpretation of the observed linewidths.
As usual, it was determined from the width of the telluric 
lines in the spectra
since their intrinsic widths are just a few km\,s$^{-1}$ and thus
negligible \citep[e.g.,][]{1982A&A...114..357B}.
Gaussian functions were fitted 
to the lines, and their mean Full Width Half Maximum (FWHM)  
are quoted in Table~\ref{logbook} as the spectral resolution,
with the error bars representing the rms fluctuations along
the field of view. 
We average over all the targets observed with the same 
instrumental configuration 
to estimate the resolution corresponding to that 
particular configuration. The weighted average for the IDS@INT 
spectra turns out to be $25.4\pm 0.6$~km\,s$^{-1}$, 
whereas in the case of  ALFOSC@NOT the measured 
resolution is $54.4\pm 3.9$~km\,s$^{-1}$.
The error bars represent the weighted rms fluctuations.
%In the case of the IDS@INT spectra,  
%it turns out to be $25\pm 5$~km\,s$^{-1}$,  where the error bar 
%represents the rms fluctuations in  
%various positions on the CCD. 
%In the case of the ALFOSC@NOT spectra,
%the measured resolution is $55\pm 20$~km\,s$^{-1}$.

%
%%%%%%%%%%%
%
%   LogFile
%
%%%%%%%%%%%
\begin{deluxetable*}{cccccccccc}
\tabletypesize{\scriptsize}
%\tablecolumns{10}
\tablewidth{0pt}%10cm}
\tablecaption{Observing Logbook}
\tablehead{
\colhead{Name\tablenotemark{a}}&
\colhead{RA~~~DEC}&	
\colhead{$t_{\rm exp}$\tablenotemark{b}}&
\colhead{Date}&
\colhead{Instrument}&
\colhead{Slit}&
\colhead{Seeing\tablenotemark{c}}&
\colhead{Pixel}&
\colhead{Scale}&
\colhead{Telluric\tablenotemark{d}}\\
&&[s]&2012&&&&&[pc/\arcsec]&[km\,s$^{-1}$]
}
\startdata
{\sc kiso3193}&08:56:08 +39:52:09&6000&Feb 10&IDS@INT&1\farcs0&1\farcs0&0.32\AA$\times$0\farcs44&124&26.0$\pm$1.9\\
{\sc kiso3867}&09:40:13 +29:35:30&6000&Feb 10&IDS@INT&1\farcs0&1\farcs0&0.32\AA$\times$0\farcs44&36&25.0$\pm$1.0\\
{\sc kiso5149}&11:16:08 +23:29:16&6000&Feb 10&IDS@INT&1\farcs0&1\farcs0&0.32\AA$\times$0\farcs44&834&28.2$\pm$1.5\\
{\sc kiso5639}&11:41:07 +32:25:37&5000&May 29&ALFOSC@NOT&0\farcs9&0\farcs8&0.51\AA$\times$0\farcs38&119&53.9$\pm$5.7\\
{\sc kiso6669}&12:31:50 +27:23:13&5000&May 29&ALFOSC@NOT&0\farcs9&0\farcs8&0.51\AA$\times$0\farcs38&299&55.6$\pm$8.7\\
{\sc kiso6877}&12:46:11 +26:15:01&5000&May 31& ALFOSC@NOT&0\farcs9&0\farcs7&
0.51\AA$\times$0\farcs38&128&54.3$\pm$6.6\\
{\sc kiso8466}&16:03:27 +19:09:46&4800&Feb 10&IDS@INT&1\farcs0&1\farcs0&0.32\AA$\times$0\farcs44&313&24.9$\pm$0.7
\enddata
\tablenotetext{a}{Named as in \citet{2012ApJ...750...95E}, Paper~I.}
\tablenotetext{b}{Total exposure time.}
\tablenotetext{c}{Mean value for the night from RoboDIMM@WHT; see
{\tt http://catserver.ing.iac.es/robodimm/}.}
\tablenotetext{d}{FWHM of the observed telluric lines, used as  proxy for spectral resolution.}
%\tablenotetext{e}{CAT stands for {\em Comit\'e de Asignaci\' on de Tiempos}, i.e.,
%Spanish time allocation committee.}
\label{logbook}
\end{deluxetable*}
%\end{deluxetable} % Just for the referee's format

\section{Determination of physical parameters}\label{determination}

Velocities,  masses, abundances and other physical parameters are determined
from the spectra. This section explains how these parameters are computed, 
including the underlying hypotheses and their uncertainties.

Bulk velocities are measured from the displacement of
H$\alpha$. We compute the displacement both as the barycenter
of the emission line, and as the center of a Gaussian function
fitted to the profile. Errors are estimated from the S/N
measured in the continuum and then propagated to the centroids
 \citep[e.g.,][Sect.~5.3]{mar71}. 
The FWHM of the profiles are also measured directly from the
profile and from the Gaussian fit. Their errors are also inferred 
from  the noise measured in the continuum by 
error propagation.

The variation of the velocity with position along the slit is 
usually referred to as the velocity curve.  The velocity curves of our targets can be 
ascribed to rotation.  In order to characterize such rotation, 
we fit an analytic RC  to the velocity curve.   
We wanted a  function that is simple but  produces 
a good match to the 
observations. The {\it universal}~RC advocated by  \citet{2007MNRAS.378...41S}
does a good job representing the observed variation.
Specifically, we adopted the dark matter
component of the universal RC given by,
\begin{equation}
U(d)=U_0+U_1\frac{d-d_0}{\sqrt{\Delta^2+(d-d_0)^2}},
\label{rc}
\end{equation}
where $U(d)$ is the velocity observed at a distance $d$, and $U_0, U_1$,
$\Delta$, and $d_0$ are free parameters to be determined 
from a non-linear fit.  The curve is simple --
$d_0$ yields a center for the RC,  $U_1$ gives its
amplitude ($U\rightarrow U_0\pm U_1$ when $d\rightarrow\pm\infty$),
and $\Delta$ provides the spatial scale for the central gradient.
We stress that the hypotheses behind 
the analytic expression~(\ref{rc})  are of little importance in our  
context. The formula is used here because it provides a good 
representation of the observations, and so provides
a smoothed version of the velocity curve. Moreover, it allows us to 
determine the center of rotation $d_0$. 
The center thus determined does not depend 
so much on the actual expression used to parameterize the RC, 
but on the fact that the curvature changes 
sign at the center of rotation, i.e.,
\begin{equation}
U^{\prime\prime}(d_0)=0,
\end{equation} 
 with $U^{\prime\prime}$ the second derivative of $U$. 
This is a general property that any anti-symmetric RC satisfies.

Dynamical masses are estimated  
from RCs and  linewidths.  Assuming the mass distribution to be 
spherically symmetric (e.g., a bulge or a dark matter halo),  
the circular velocity $U_c$ that
balances the gravitational pull depends only on the 
mass enclosed within the radius $\rho$,  $M(\rho)$,    
\begin{equation}
U_c^2(\rho)=G\, M(\rho)/\rho,
\end{equation} 
with $G$ the gravitational constant.
We assume that the measured macroscopic velocities 
are the circular velocities affected by the inclination $i$
of the galaxy plane, i.e., $U=U_c\,\sin i$. We also assume the 
spectrograph slit to be oriented  along the galaxy  major axis,
implying $\rho=d-d_0$. Then the inner mass of the galaxy up to
distance $d$ has the usual expression,
\begin{equation} 
M(d)\,\sin^2 i =(d-d_0)\,U^2(d)/G,
\label{mass_rot1}
\end{equation} 
which using astrophysical units turns out to be  
\begin{equation}
M(d)\,\sin^2  i=(2.33\cdot10^{5}{\rm M_\odot})~(d-d_0)\,U(d)^2,
\label{mass_rot2}
\end{equation}
with $d$ measured in kpc and U in km\,s$^{-1}$.
Equation~(\ref{mass_rot2})  is an approximation;  however, including
more realism in the mass distribution 
(e.g., pressure support, 
\citeauthor{2010ApJ...721..547D}~\citeyear{2010ApJ...721..547D};
or non-spherical components, 
\citeauthor{1997ASPC..117....1S}~\citeyear{1997ASPC..117....1S})
would only modify the scaling factor as a correction of order one.
Masses of individual clumps are inferred from linewidths 
assuming virial equilibrium. If the isotropic velocity distribution 
that balances the clump gravity has a dispersion $\sigma$, then
\citep[][Sect.~4.2]{2009AJ....137.3437B}
\begin{equation}
M=(1.20\cdot10^{5}{\rm M_\odot})~R_e\,{\rm FWHM}^2,
\label{veloc_turb}
\end{equation}
where
\begin{equation}
{\rm FWHM}=2\sqrt{2\ln 2}\,\sigma.
\label{sigma2fwhm}
\end{equation}
The symbol $R_e$ stands for the half-light radius of the clump, so
that  Eq.~(\ref{veloc_turb}) is equivalent to Eq.~(\ref{mass_rot2}) after 
including the appropriate scaling factors. As in the 
case of  Eq.~(\ref{mass_rot2}), distances are measured in 
kpc and linewidths in km\,s$^{-1}$.
The linewidth in Eq.~(\ref{veloc_turb})
is not the observed width ${\rm FWHM_o}$, but the 
width corrected for instrumental spectral resolution  
${\rm FWHM_i}$, thermal motions in the nebula 
${\rm FWHM_t}$, and natural width of H$\alpha$
${\rm FWHM_n}$, 
\begin{equation}
{\rm FWHM^2 = FWHM_o^2 - FWHM_i^2-FWHM_n^2-FWHM_t^2}, 
\label{correct_fwhm}
\end{equation}
see, e.g., 
\citet[][]{1981MNRAS.195..839T,1999MNRAS.302..677M}.
We use the measured widths of the telluric lines as proxy
for instrumental broadening (Table~\ref{logbook}).
The thermal broadening is assumed be the same for all galaxies 
at all positions. Its value has been set to a 
representative round number FWHM$_{\rm t}= 25$\,km\,s$^{-1}$, which 
approximately corresponds to H atoms at 14000\,K, a temperature 
typical of HII regions. 
The actual FWHM$_{\rm t}$ is of secondary 
importance since 
the range of possible values 
is significantly smaller than the FWHM
resulting from Eq.~(\ref{correct_fwhm})\footnote{
FWHM$_{\rm t}$ varies only from 19 to and 26\,km\,s$^{-1}$
for temperatures between 8000 and 15000\,K.
Moreover, we did the exercise of calculating masses
also with the extreme value of  FWHM$_{\rm t}$=0,
to check that this assumption does not modify the conclusions 
in the work. Setting  FWHM$_{\rm t}$=0 is
equivalent to including thermal motions  as part of 
the virial equilibrium represented by Eq.~(\ref{veloc_turb}). 
The exercise implies
%It was  checked by calculating masses also with
%Incidentally, this implies 
that we do not have to 
worry about whether the thermal motions contribute or not 
to the virial equilibrium.% represented by Eq.~(\ref{veloc_turb}). 
%Incidentally, it implies that 
Including or not  thermal motions as part of the kinetic energy 
does not significantly modify the masses estimated in the work.
}. 
Finally, the natural width of H$\alpha$ is of 
the order of 7\,km\,s$^{-1}$ \citep[e.g.,][]{2006A&A...455..539R}.
The radius $R_e$ used to estimate dynamical masses from line
widths using Eq.~(\ref{veloc_turb}) is computed fitting a 
1-D Gaussian to the light distribution across the galaxy head. The observed 
light profile is assumed to represent a 1-D cut across the center of a 2D-Gaussian, 
which readily provides the half-light radius from the width of the fitted Gaussian
-- in a 2D Gaussian, the FWHM is twice the half-light radius. 
The observed radii are corrected for seeing using a formula similar to 
Eq.~(\ref{correct_fwhm}),
namely,
\begin{equation}
R_e^2=R_{e0}^2-(S/2)^2,
\end{equation}
where $R_{e0}$ stands the measured effective radius and 
$S$ represents the seeing, i.e.,  the FWHM of the seeing disk
as given in Table~\ref{logbook}.

%%%
The metallicity of the gas is commonly inferred by combining emission-line 
fluxes of several atomic species to derive 
their relative abundances  \citep[e.g.,][]{1981ARA&A..19...77P,1989agna.book.....O}. 
This approach is the so-called direct method or temperature-based method,
and it is to be preferred whenever possible. However, it involves measuring fluxes 
of lines spread throughout the visible-IR spectrum, so is 
expensive observationally. Fortunately, we have  alternatives 
called strong-line methods 
\citep[e.g.,][]{2005A&A...437..849S,2008ApJ...681.1183K}, where 
the
metallicity is estimated  empirically  by relating the ratio of a 
few selected line fluxes with the abundance of  a particularly 
relevant metal (typically oxygen).
The one proposed by \citet{2002MNRAS.330...69D}  
turns out to be ideal in our case, when only the spectral region 
around H$\alpha$ is available.  It yields the oxygen abundance from the ratio 
of [NII]$\lambda$6583 to H$\alpha$, and [NII]$\lambda$6583
automatically appears in the spectra next to H$\alpha$
(see Fig.~\ref{reduction}). 
We use the calibration by  \citet{2009MNRAS.398..949P},
\begin{equation}
12+\log({\rm O/H})=9.07+0.79 \log({\rm [NII]}\lambda 6583/{\rm H}\alpha),
%12+\log({\rm O/H})=8.9+0.57 \log({\rm [NII]}\lambda 6583/{\rm H}\alpha).
\label{metal_equation}
\end{equation}
particularly suited for low metallicity targets
\citep[c.f.,][]{2004MNRAS.348L..59P}.
Equation~(\ref{metal_equation}) provides the O metallicity 
from the flux in a N line, therefore, it may be biased 
in objects having unusual N/O.
In order to discard this potential bias in our O metallicities, 
we also  estimate the ratio of N to O using
%The 
%ratio of nitrogen to oxygen can be estimated from 
the sulfur lines [SII]$\lambda$6717 and [SII]$\lambda$6731  present in 
most of our spectra.
We use the calibration
\begin{equation}
\log({\rm N/O})=1.26 \log({\rm [NII]}\lambda 6583/{\rm [SII]}\lambda 6717,6731)-0.86,
\label{noequation}
\end{equation}
also by \citet{2009MNRAS.398..949P}.

The {\em spectral flux} of the galaxy is computed by  integrating the 
observed spectra between their two extreme wavelengths  
(from 6242\,\AA\ to 6935\,\AA\  for INT spectra, and 
from 6368\,\AA\ to 6840\,\AA\ for NOT spectra).
The flux in H$\alpha$ results from integration of 
the emission line profile around its maximum, once the 
underlying continuum was removed. The integration includes
a 10\,\AA\ wide region around the peak emission,
whereas the continuum was obtained by 
fitting a linear function to two continuum windows, 10\,\AA\, wide, 
outside the line.  The H$\alpha$ equivalent width (EW) is 
inferred from the H$\alpha$ flux dividing by this continuum. 
Continuum and emission lines are combined in the {\em spectral flux}
for simplicity, and it suffices to compare the limited extension 
of the emission line region inferred from H$\alpha$  
with the rest of the galaxy. 

%%%%%%%%%%%%%%%%%

\section{Velocity curves and linewidths}\label{dynamical_properties}

\begin{figure*}
\includegraphics[width=\textwidth]{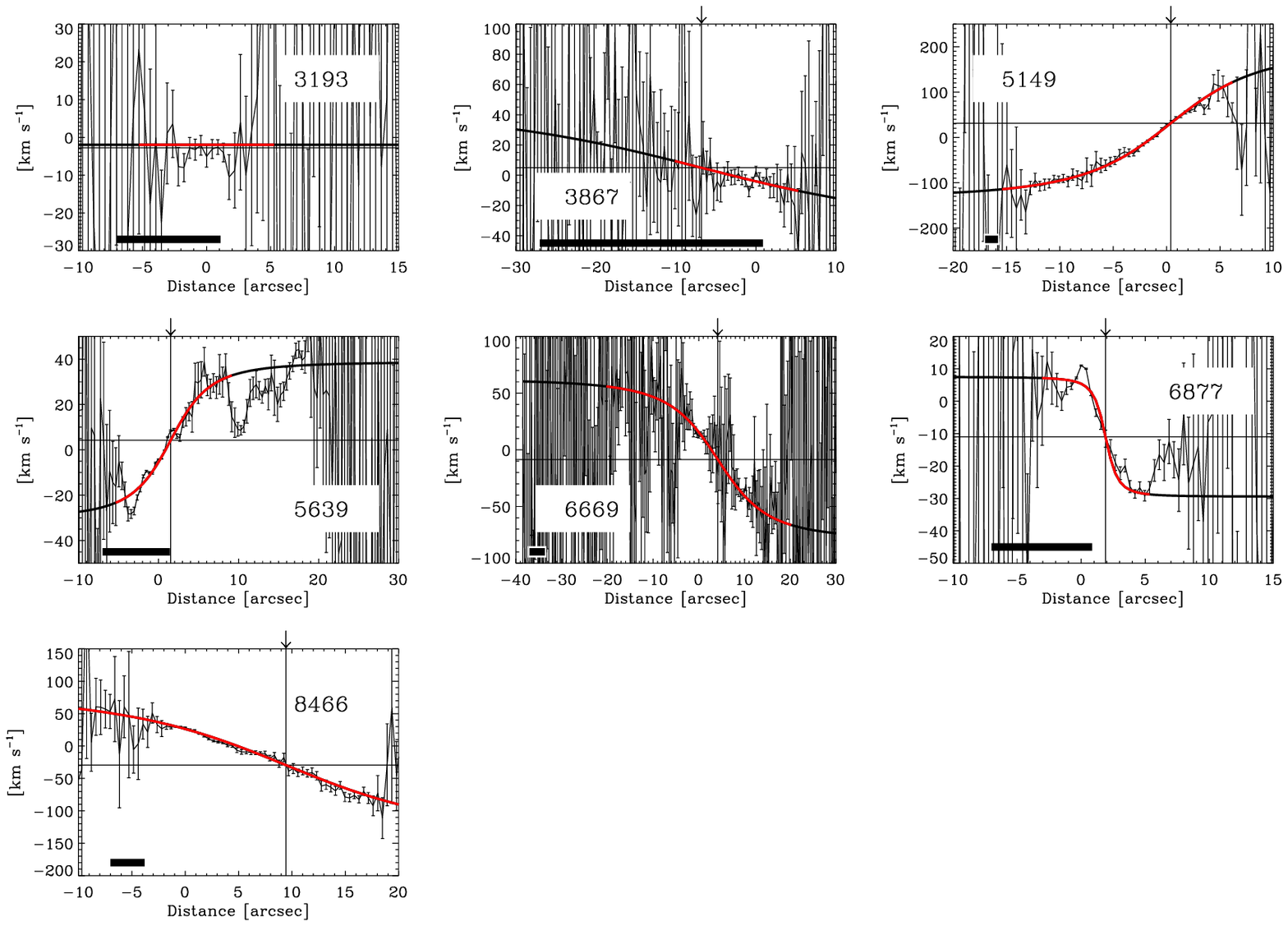}
\caption{
Velocity curves of the seven tadpole galaxies.
The abscissae represent distances along the major axes of the galaxies 
from the position of the tadpole head; i.e., the brightest point
on the galaxy. The range of distances differs for the different
targets, but the horizontal bar in each panel gives a common length 
scale corresponding to 1\,kpc.
The points with error bars show the observations whereas the thick 
solid line represents the best fit of the observed points to the 
analytic RC.
The part of the RC shown in red indicates the 
portion of the velocity curve used for fitting.
The thin horizontal and vertical lines 
indicate the systemic velocity and the center of rotation
obtained from the fit, respectively.
The little arrows on top of each panel also indicates
the center of rotation.
The zero of the velocity scale 
is set by velocity of the spatially integrated
spectrum of the galaxy.  Positive velocities are redshifts.
The ordering of the galaxies is identical 
to that in Fig.~\ref{images}.
}
\label{velocity_curves}
\end{figure*}

Figure~\ref{velocity_curves} shows the velocity curves of the tadpole
galaxies included in our study. Five out of the seven targets 
show velocity gradients interpreted as rotation. 
The figure includes the best fit to  the analytic RC in Eq.~(\ref{rc}),
which  does a good job reproducing the observations.
We use for fitting the portion of the velocity curve indicated
in red in Fig.~\ref{velocity_curves}, which includes all positions
but the extremes with large error bars or obvious distortions.
Sometimes the interpretation of the velocity curve
as a RC is obvious (e.g., {\sc kiso8466}), but other times 
the curve looks more like a perturbed RC (e.g., {\sc kiso5639}). 
 {\sc kiso3193} and {\sc kiso3867}
have a rather flat velocity curve, and therefore 
no obvious rotation. However, one of them, 
{\sc kiso3867}, shows a systematic line shift of the order 10--20~km\,s$^{-1}$ 
between the two extremes of the galaxy (Fig.~\ref{velocity_curves}). 
The amplitude is of the order of the error bars, but the displacement
is in the raw data as judged by inspection of the individual H$\alpha$ profiles.
%Such  velocity gradient may be caused by a mild solid-body rotation. 
Table~\ref{obs_summary}
contains the amplitudes of the RCs as assigned by the fit (i.e.,
$U_1$ in Eq.~[\ref{rc}]). 
Figure~\ref{velocity_curves} represents velocities obtained from the 
barycenter of H$\alpha$ (Sect.~\ref{determination}). The velocities from the  
Gaussian fit are not shown because the two estimates differ 
only by a few km\,s$^{-1}$, a difference  always smaller 
than the error bar assigned to each velocity measurement.

%%%%%%%%%%%%%%%%%%%
%
%   Parameters of individual galaxies
%
%%%%%%%%%%%%%%%%%%%
%\renewcommand{\arraystretch}{0.3}
%%%%%%%%%%%%%
\begin{deluxetable*}{lccccccccc}
\tabletypesize{\scriptsize}
%\tablecolumns12}
\tablewidth{19cm}%0pt}%12cm}
\tablecaption{Physical parameters}
\tablehead{
\colhead{Name\tablenotemark{a}  \hfill Distance\tablenotemark{a}}&
%\colhead{Distance}&
\colhead{FWHM\tablenotemark{b}}&
\colhead{$\log M_h$ D\tablenotemark{c}}&
\colhead{$R_e$\tablenotemark{d}}&
\colhead{$\log M_h$ Ph\tablenotemark{e}}&
\colhead{$U_1$\tablenotemark{f}}&
\colhead{$d_0$\tablenotemark{g}}&
\colhead{$\log M$ D\tablenotemark{h}}&
\colhead{$\log M$ Ph\tablenotemark{i}}&
\colhead{\hspace{-0.2cm}12+log(O/H)\tablenotemark{j}}\\
\hfill [Mpc]~~~
&[km\,s$^{-1}$]&[$M_\odot$]&[kpc]&[$M_\odot$]&[km\,s$^{-1}$]&[kpc]&[$M_\odot$]&[$M_\odot$]
}
\startdata
% table before the 1st referee's report
%
%{\sc kiso3193}\hfill 25.5~~~&$39.6\pm 0.4$&$7.61\pm0.02$&$0.221\pm0.009$&$6.4\pm0.1$&\nodata&\nodata&\nodata&$7.3\pm0.4$&$7.89\pm0.31$\\
%
%{\sc kiso3867}\hfill 7.4~~~&$35.7\pm 0.4$&$7.22\pm 0.01$&$0.109\pm0.003$&$5.2\pm0.3$&\nodata&\nodata&\nodata&$6.8\pm0.4$&$8.03\pm0.26$\\
%{\sc kiso5149}\hfill 172\phd\phn~~~&$70.2\pm 0.9$&$9.03\pm0.02$&$1.825\pm0.064$&$8.9\pm0.1$&$168\pm92$&$0.3\pm3.0$&$10.8\pm 0.6$&$9.6\pm0.4$&$8.42\pm0.06$\\
%
%{\sc kiso5639}\hfill 24.5~~~&$45.8\pm 0.5$&$7.21\pm0.03$&$0.065\pm0.004$&$6.7\pm0.2$&$ 34.7\pm6.2$&$0.180\pm0.064$&$8.2\pm 0.2$&$7.7\pm0.4$&$7.48\pm0.04$\\
%
%{\sc kiso6669}\hfill 61.6~~~&$46.0\pm 0.3$&$7.72\pm0.02$&$0.206\pm0.011$&$7.0\pm0.1$&$71.4\pm4.6$&$1.24\pm0.16$&$9.8\pm 0.1$&$8.9\pm 0.4$&$8.11\pm0.12$\\
%
%{\sc kiso6877}\hfill 26.3~~~&$24.8\pm0.1$&$6.31\pm0.06$&$0.028\pm 0.004$&$6.3\pm 0.1$&$19.0\pm 1.7$&$0.246\pm0.020$&$7.7\pm 0.1$&$7.3\pm0.4$&$7.40\pm 0.09$\\
%
%{\sc kiso6877}\tablenotemark{i}\hfill26.3~~~&$24.3\pm 0.1$&$6.43\pm0.04$&$0.038\pm 0.003$&$6.3\pm 0.1$&$19.0\pm 1.6$&$0.242\pm0.020$&$7.6\pm 0.1$&$7.3\pm0.4$&$7.41\pm 0.07$\\
%
%{\sc kiso8466}\hfill 64.5~~~&$61.2\pm0.8$&$8.19\pm 0.03$&$0.345\pm0.021$&$8.1\pm0.1$&$117\pm16$&$2.95\pm0.36$&$10.0\pm 0.1$&$9.2\pm0.3$&$8.14\pm0.05$
%
% table with the new error bars
{\sc kiso3193}\hfill 25.5~~~~~&$39.3\pm 2.0$&$7.61\pm0.04$&$0.221\pm0.009$&$6.4\pm0.1$&\nodata&\nodata&\nodata&$7.3\pm0.4$&$7.89\pm0.31$\\
{\sc kiso3867}\hfill 7.4~~~~~&$35.4\pm 2.2$&$7.22\pm 0.04$&$0.109\pm0.003$&$5.2\pm0.3$&\nodata&\nodata&\nodata&$6.8\pm0.4$&$8.03\pm0.26$\\
{\sc kiso5149}\hfill 172\phd\phn~~~~~&$70.1\pm 3.0$&$9.03\pm0.03$&$1.825\pm0.064$&$8.9\pm0.1$&$168\pm92$&$0.3\pm3.0$&$10.8\pm 0.6$&$9.6\pm0.4$&$8.42\pm0.06$\\
{\sc kiso5639}\hfill 24.5~~~~~&$46\phd\phn\pm 11\phd$&$7.22\pm0.15$&$0.065\pm0.004$&$6.7\pm0.2$&$ 34.7\pm6.2$&$0.180\pm0.064$&$8.2\pm 0.2$&$7.7\pm0.4$&$7.48\pm0.04$\\
{\sc kiso6669}\hfill 61.6~~~~~&$47\phd\phn\pm 11\phd$&$7.73\pm0.14$&$0.206\pm0.011$&$7.0\pm0.1\phd$&$71.4\pm4.6$&$1.24\pm0.16$&$9.8\pm 0.1$&$8.9\pm 0.4$&$8.11\pm0.12$\\
{\sc kiso6877}\hfill 26.3~~~~~&$22\phd\phn\pm 23\phd$&$6.19\pm0.66$&$0.028\pm 0.004$&$6.3\pm 0.1$&$19.0\pm 1.7$&$0.246\pm0.020$&$7.7\pm 0.1$&$7.3\pm0.4$&$7.40\pm 0.09$\\
{\sc kiso6877}\tablenotemark{k}\hfill26.3~~~~~&$21\phd\phn\pm 24\phd$&$6.30\pm0.70$&$0.038\pm 0.003$&$6.3\pm 0.1$&$19.0\pm 1.6$&$0.242\pm0.020$&$7.6\pm 0.1$&$7.3\pm0.4$&$7.41\pm 0.07$\\
{\sc kiso8466}\hfill 64.5~~~~~&$61.0\pm2.9$&$8.19\pm 0.04$&$0.345\pm0.021$&$8.1\pm0.1$&$117\pm16$&$2.95\pm0.36$&$10.0\pm 0.1$&$9.2\pm0.3$&$8.14\pm0.05$
\enddata
\tablenotetext{a}{From \citet{2012ApJ...750...95E}, Paper~I.}
\tablenotetext{b}{H$\alpha$ width at the tadpole head.}
\tablenotetext{c}{Dynamical mass of the head -- from Eq.~(\ref{veloc_turb}).}
\tablenotetext{d}{Half-light radius of the head.}
\tablenotetext{e}{Photometric mass of the head, from Paper~I, Table~3.}
\tablenotetext{f}{Amplitude of the rotation curve -- see Eq.~(\ref{rc}).}
\tablenotetext{g}{Center of rotation relative to the tadpole head -- see Eq.~(\ref{rc}).}
\tablenotetext{h}{Dynamical mass of the galaxy  -- see text for its computation.}
\tablenotetext{i}{Photometric mass of the galaxy, from Paper~I, Table~3.}
\tablenotetext{j}{Metallicity at the head -- from Eq.~(\ref{metal_equation}).}
\tablenotetext{k}{Second independent reduction.}
\label{obs_summary}
\end{deluxetable*}
%%%%%%%%%%%%%%

\begin{figure*}
\includegraphics[width=\textwidth]{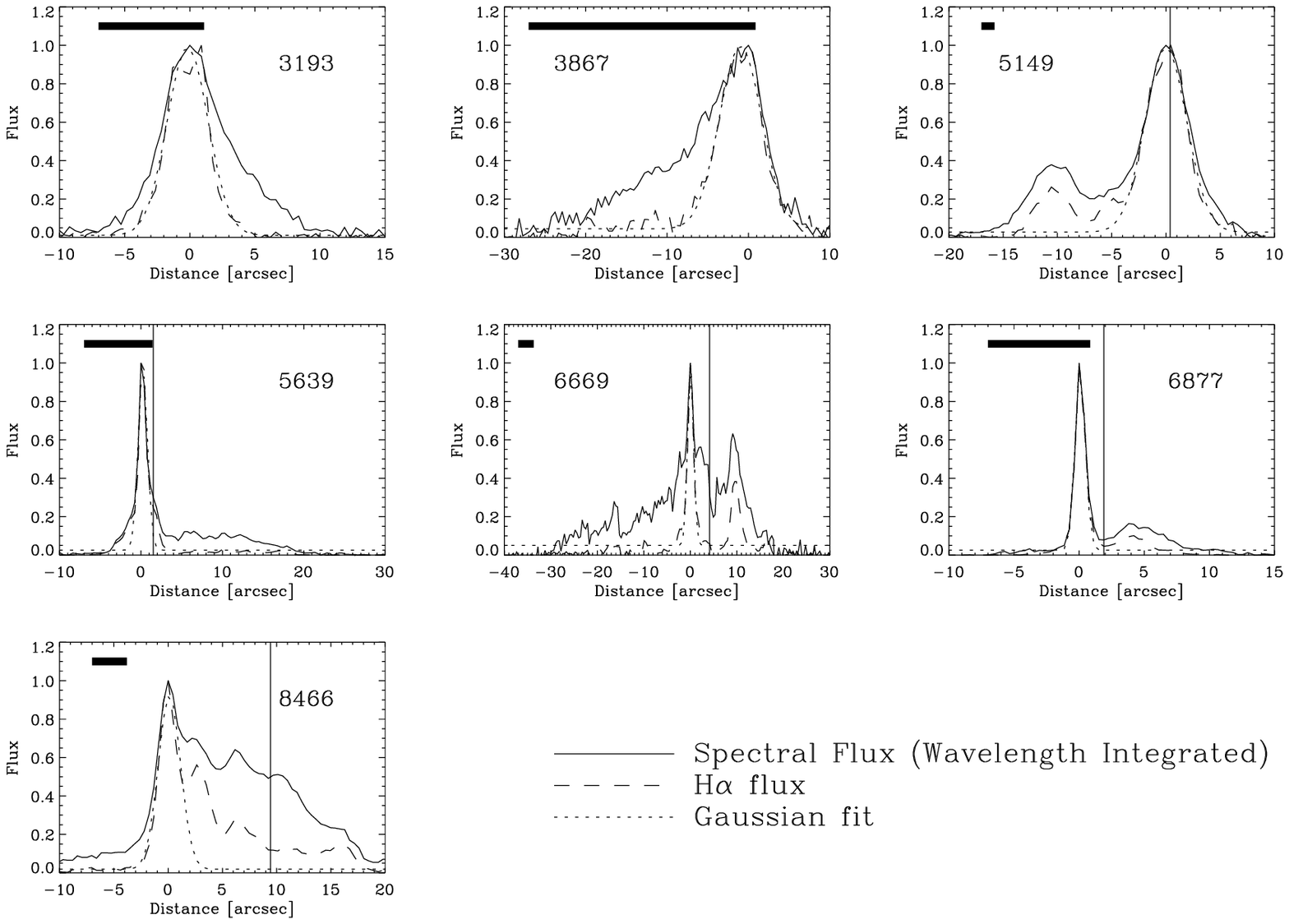}
\caption{
Spectral flux (the solid lines) and H$\alpha$ flux (the dashed lines).
Note the obvious lopsidedness of the light
distributions, as expected in tadpole galaxies.
The origin of distances has been set as the maximum of 
the spectral flux distribution. 
The dotted line represents a Gaussian fitted to the 
H$\alpha$ flux around the head.
The horizontal bar in each panel gives a common length 
scale corresponding to 1\,kpc.
Formal error bars for photometry are not included 
since they are negligible small.
The thin vertical solid lines indicate the center of rotation
obtained from the RC fit shown in Fig.~\ref{velocity_curves}.
The ordering of the galaxies and the range of abscissae are identical 
to those in Fig.~\ref{velocity_curves}.
}
\label{brightness}
\end{figure*}

Table~\ref{obs_summary} also contains the center of rotation $d_0$, which
often differs from zero, i.e., from the position of the tadpole head.  
As argued in Sect.~\ref{determination}, its estimate is
fairly robust since it comes from the point where the curvature
of the RC changes sign -- accordingly, it has
the small formal error bars provided in Table~\ref{obs_summary}. 
Figure~\ref{velocity_curves} shows that three  out of the five rotating galaxies, 
explicitly {\sc kiso6669},  {\sc kiso6877} and {\sc kiso8466},
have their center of rotation displaced with respect to the head 
by more than the 1\arcsec\ uncertainty 
introduced by seeing 
(see Table~\ref{logbook}).
Considering that the rotation center 
points out the center of the galaxy, our results indicate that 
often the  star-forming region at  the head is displaced with 
respect to the center  of the galaxy. 
%added below
This suggests that the head is not a bulge-like central spheroid.
The fact that the center of rotation is sometimes displaced
from the heads may be easier to appreciate in Fig.~\ref{brightness}.

%%%%%%%%%%%%%%%%%
%%%%%%%%%%%%%%%
 
%%%%%%%%%%

\begin{figure*}
\includegraphics[width=\textwidth]{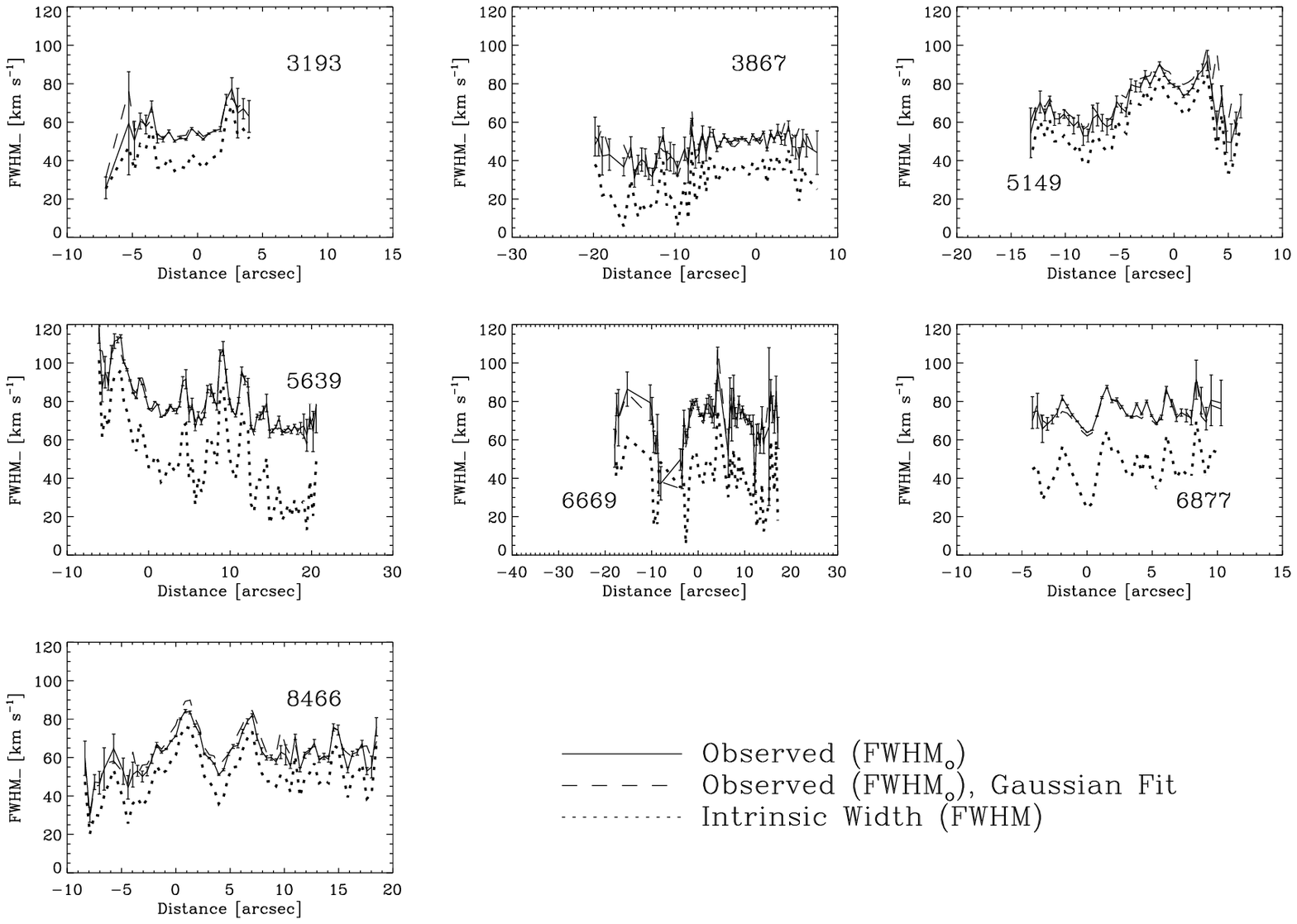}
\caption{
FWHM of H$\alpha$ inferred directly from the
emission line profile (the solid lines), from the Gaussian fit 
(the dashed lines), and after correction for instrumental, thermal
and natural
broadenings (the dotted lines).
Formal error bars for the direct measurement are represented too.
The ordering of the panels and the abscissae are identical 
to those in Fig.~\ref{velocity_curves}.
}
\label{linewidths}
\end{figure*}

\begin{figure}
\includegraphics[width=0.5\textwidth]{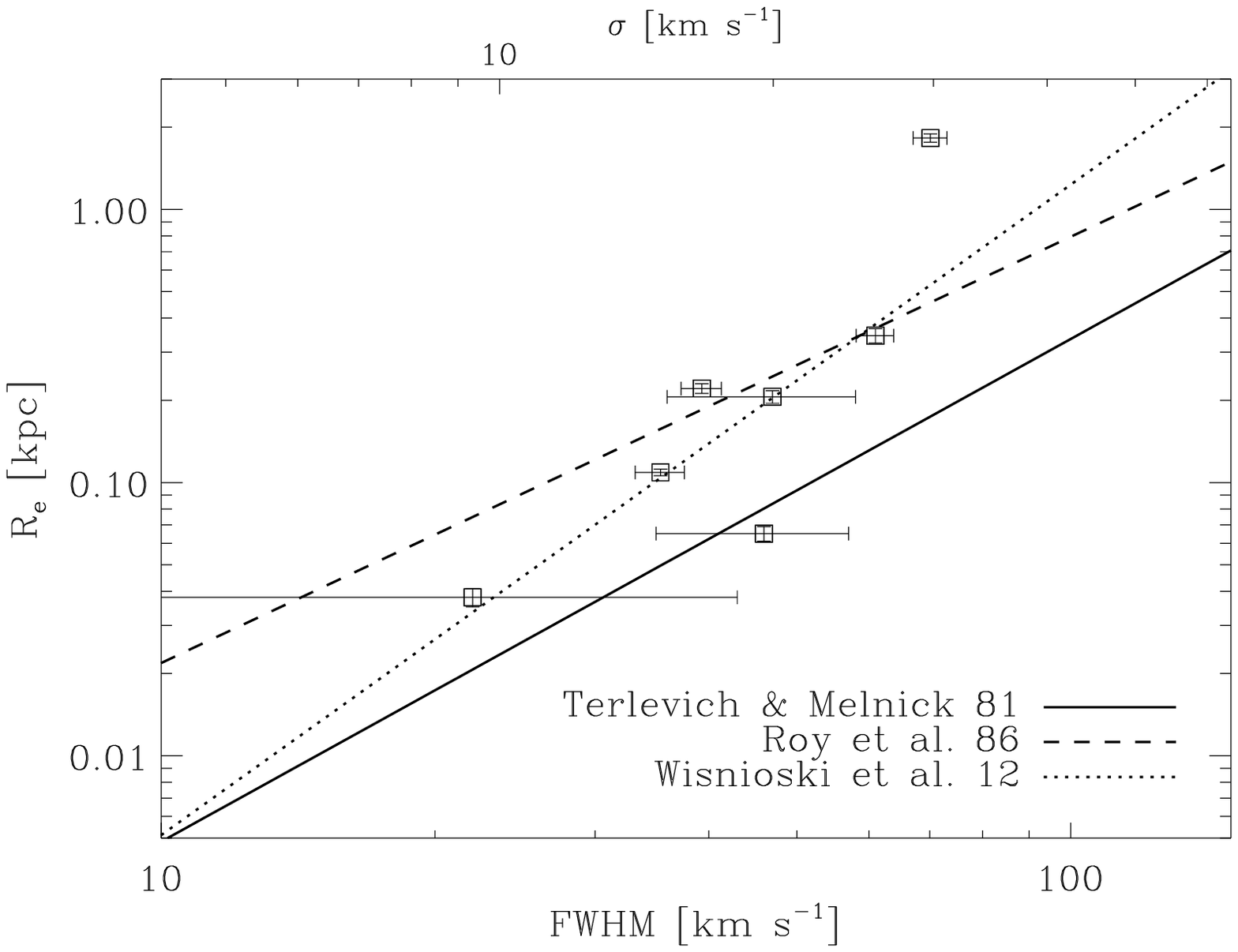}
\caption{
Half-light radius of the tadpole head versus H$\alpha$ 
linewidth. The velocity dispersions are usually
parameterized in terms of $\sigma ={\rm FWHM}/(2\sqrt{2\ln 2})$,
and its scale is given in the abscissae on top of the figure.  
The observed points follow the trends observed in 
giant HII regions -- 
solid,  dashed, and the dotted lines  
correspond to the empirical calibration of the 
relationship by \citet{1981MNRAS.195..839T}, \citet{1986ApJ...300..624R}, 
and \citet{2012MNRAS.422.3339W},
respectively. 
%
%Formal error bars are not included since they have
%sizes similar to the symbols.
}
\label{sigmare}
\end{figure}

Figure~\ref{linewidths} shows the FWHM of H$\alpha$ as inferred directly from the
emission line profile (the solid lines) and from the Gaussian fit (the dashed lines).
%They are in good agreement. 
A significant part of the observed linewidths is
due to instrumental broadening. The 
widths corrected for instrumental effects and thermal motions are given as 
dotted lines in the figure.
We use the directly inferred widths to estimate the intrinsic 
widths, but using them or the widths from the Gaussian fit 
render similar results since the two measurements are in very good 
agreement (Fig.~\ref{linewidths}).
First note that the widths of all galaxies, large and small, are
of the order of 
20--70\, km\,s$^{-1}$. 
These widths are typical
of giant HII regions such as 30 Doradus
rather than the widths of the HII regions observed 
in large spirals \citep[e.g.,][]{1988A&A...198..283O,1994vsf..book...25M}. 
Figure~\ref{sigmare} shows the variation of the 
tadpole head size as a function of its velocity dispersion. 
The two quantities are known to be correlated in giant HII regions, 
so that the larger the dispersion the larger the size 
\citep[e.g.,][]{1994vsf..book...25M,2000AJ....120..752F}. We %2000AJ....119.2166F}.
find the tadpole heads to follow such trend as characterized 
by \citet{1981MNRAS.195..839T},
\citet{1986ApJ...300..624R},
or more recently by 
\citet{2012MNRAS.422.3339W}
(see Fig.~\ref{sigmare}).    
In the case of {\sc kiso6877}, the error bars of the line width 
at the head are so large that the width is actually an 
upper limit,  but even with this caveat in mind,
the measurement is consistent with its head 
being a giant HII region (Fig.~\ref{sigmare}).

The FWHM also varies along the galaxies, and the
fluctuations are correlated neither with the spectral flux 
nor with the H$\alpha$ flux (c.f. Fig.~\ref{brightness} and \ref{linewidths}).   
If anything, there is a tendency for the tadpole heads to 
coincide with local minima of linewidth
(see  {\sc kiso5149}, {\sc kiso5639}, and 
{\sc kiso6877} in Fig.~\ref{linewidths}). 
Similarly, the center of rotation of the rotating
galaxies does not seem to be associated with
extremes of the FWHM curve.
%%%%%%%%
Interpreting the decrease of linewidth associated
with the head is not straightforward. One may 
try to explain its origin
in the context of forming stars 
in a highly turbulent galaxy. 
Only where the turbulence 
is low enough the conditions that trigger star formation 
are met, and this preference for low turbulence regions
is what we detect.
%\citep[Bruce][references needed]{?}.
%\modified{
%(From this point on, the paragraph could be removed.)
%}
%One may also explain it as the presence of a
%massive clump that extends even beyond the observed 
%tadpole head. 
%Then the (gravitationally driven) motions increase
%with distance, giving rise to the increase of line
%widths\footnote{It is similar the rotation curves of galaxies,
%whose velocity increases outwards as more mass contributes 
%to the gravitational pull.}.
%A third possibility is that the motions outside the cluster
%are generated not
%by gravity but by turbulence produced by the winds and the radiation field 
%from the stellar cluster. Then the turbulent velocities may increase
%with distance from the stellar cluster, as  happens with this and this other 
%example from the literature \citep[e.g., give examples][Casiana?]{?,?}. 

The square of the ratio between velocity dispersion and  
rotation gives the Jeans length relative 
to the galaxy size 
\citep[e.g.,][]{2009ApJ...694L.158B}.  
In other words, it provides the relative size of the clumps 
to be produced by gravitational instability.   
Using  Eq.~(\ref{sigma2fwhm}) and the data in Table~\ref{obs_summary},
the ratio dispersion to rotation 
$\sigma/U_1$ turns out  to be between 0.2 and 0.6 at the 
tadpole heads, which yields expected clump sizes between 
0.03\,\%  and 30\,\% galaxy radii. The relative size 
increases with decreasing galaxy mass, and it 
may be a coincidence, but the largest ratios correspond 
to the two XMP galaxies in the sample (see Sect.~\ref{metalcont}). 
In some cases the predicted clumps have sizes comparable
to those of the observed heads. The ratio  $\sigma/U_1$
also measures whether the object is supported by random 
motions or rotation. 
Our disks have ratios similar to the turbulence-supported
clumpy galaxies observed at high redshift 
\citep[e.g.,][]{2012MNRAS.420.3490C}.  

%A final caveat  is order. We measure motions of the ionized gas, which do not 
%necessarily trace stellar motions. However, we use them 
%to estimate dynamical masses, which implicitly 
%assumes the velocities to trace collisionless test particles moving in 
%the gravitational potential, i.e., to trace stars. Even though using gas 
%motions as proxy for stellar motions is only an 
%approximation,  it suffices in 
%our case. \citet[][]{2008MNRAS.387.1099P} compare the rotation from gas
%and stars in a number of galaxies and even though the 
%gas kinematics turns out to be more irregular than the star kinematics,
%the global trends are similar and the differences
%smaller than 10\,\%. Similarly,  \citet{2012A&A...546A..52D}
%study RCs from model galaxies resulting from realistic numerical 
%simulations, and they find the gas-inferred RCs to follow the true RCs. 
%(Jairo: would you accept these statements?)
%\casiana{she suggests to remove the full paragraph.}

%%%%%%%%%%%%%%%%%%%%
\section{Fluxes and equivalent widths}\label{fluxes_and_else}

Figure~\ref{brightness} shows the variation across
the galaxies of the spectral flux and the H$\alpha$ flux
(the solid lines and the dashed lines, respectively).
The position where the two fluxes are largest coincides.
(As we already pointed out, the origin of distances used along the 
paper has been set at  the maximum of the spectral flux 
distribution.) Note also the obvious lopsidedness of all light
curves, as expected in tadpole galaxies.
Another important feature is the extension of the H$\alpha$
emission as compared to the spectral flux emission which includes
H$\alpha$ plus continuum. H$\alpha$ is more
concentrated, implying that the region producing the total
emission extends further away from  the star-forming regions.
In other words, it shows the existence of an underlying 
galaxy, with old stellar populations 
that contribute to the spectral flux but not to H$\alpha$.

The dotted line in Fig.~\ref{brightness} represents a Gaussian 
plus a constant term fitted to  the H$\alpha$ flux around the head. 
The fits are good, and they allow us to assign a 
radius to the head. Table~\ref{obs_summary} lists the effective 
radii of the heads $R_e$, defined as the radius enclosing half
of their light (see Sect.~\ref{determination}).   They span a wide range of values 
from 1.8\,kpc to 50\,pc, reflecting the wide range of intrinsic
galaxy sizes.  
%

%%%%%%%%%%%%%%%%%
\begin{figure*}
\includegraphics[width=\textwidth]{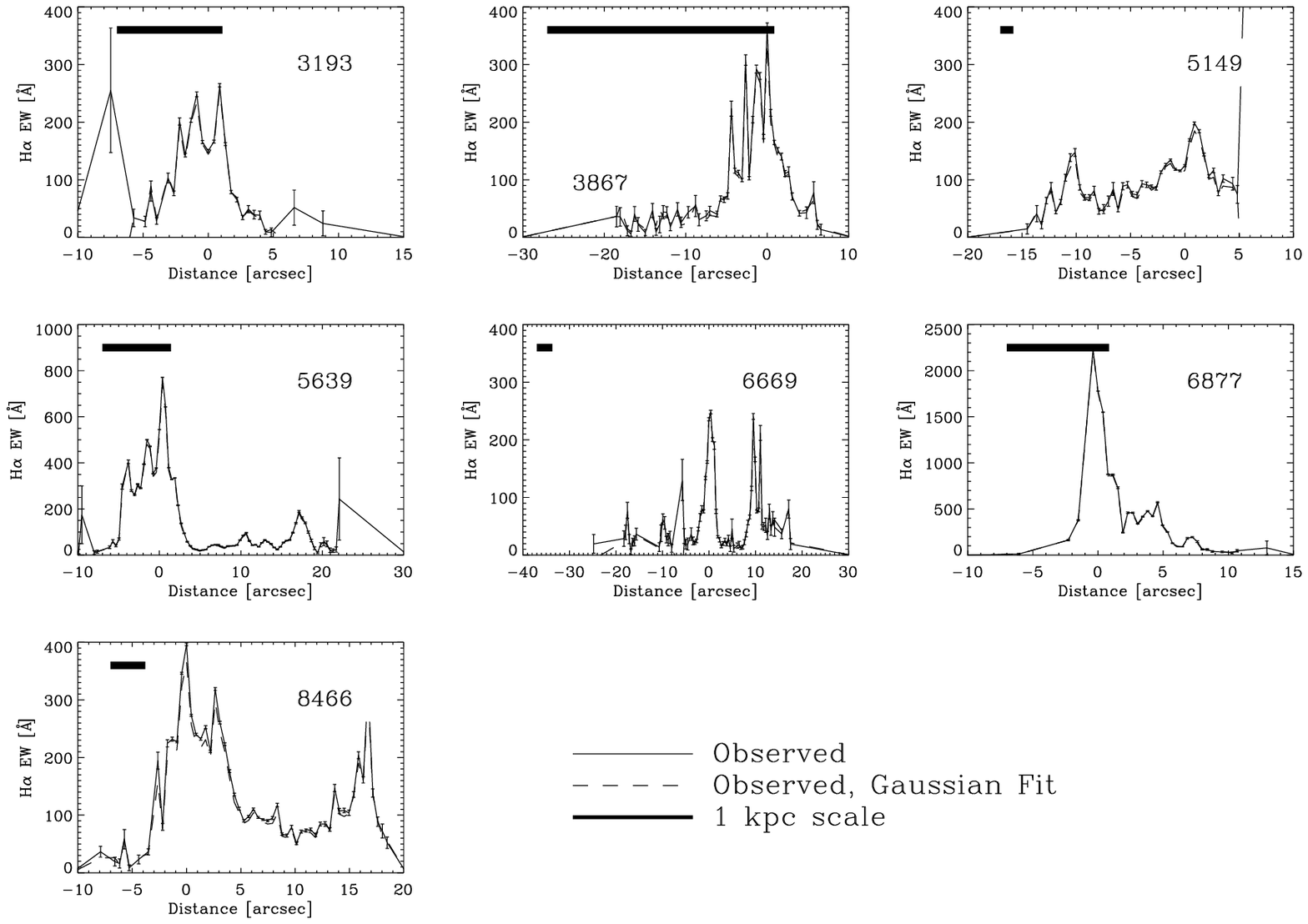}
\caption{
H$\alpha$ EW in \AA . 
EWs obtained directly from the line profile (the solid lines)
and  through a Gaussian fit (the dashed lines) 
are included, although they are difficult to distinguish.
The range of distances differs for the different
targets, but the horizontal bar in each panel gives a common length 
scale corresponding to 1\,kpc.
Only points where the EW is larger than its formal error bar 
are included.
The ordering of the galaxies and the abscissae are identical 
to those in Fig.~\ref{velocity_curves}.
}
\label{equivalentw}
\end{figure*}
\begin{figure}
\includegraphics[width=0.45\textwidth]{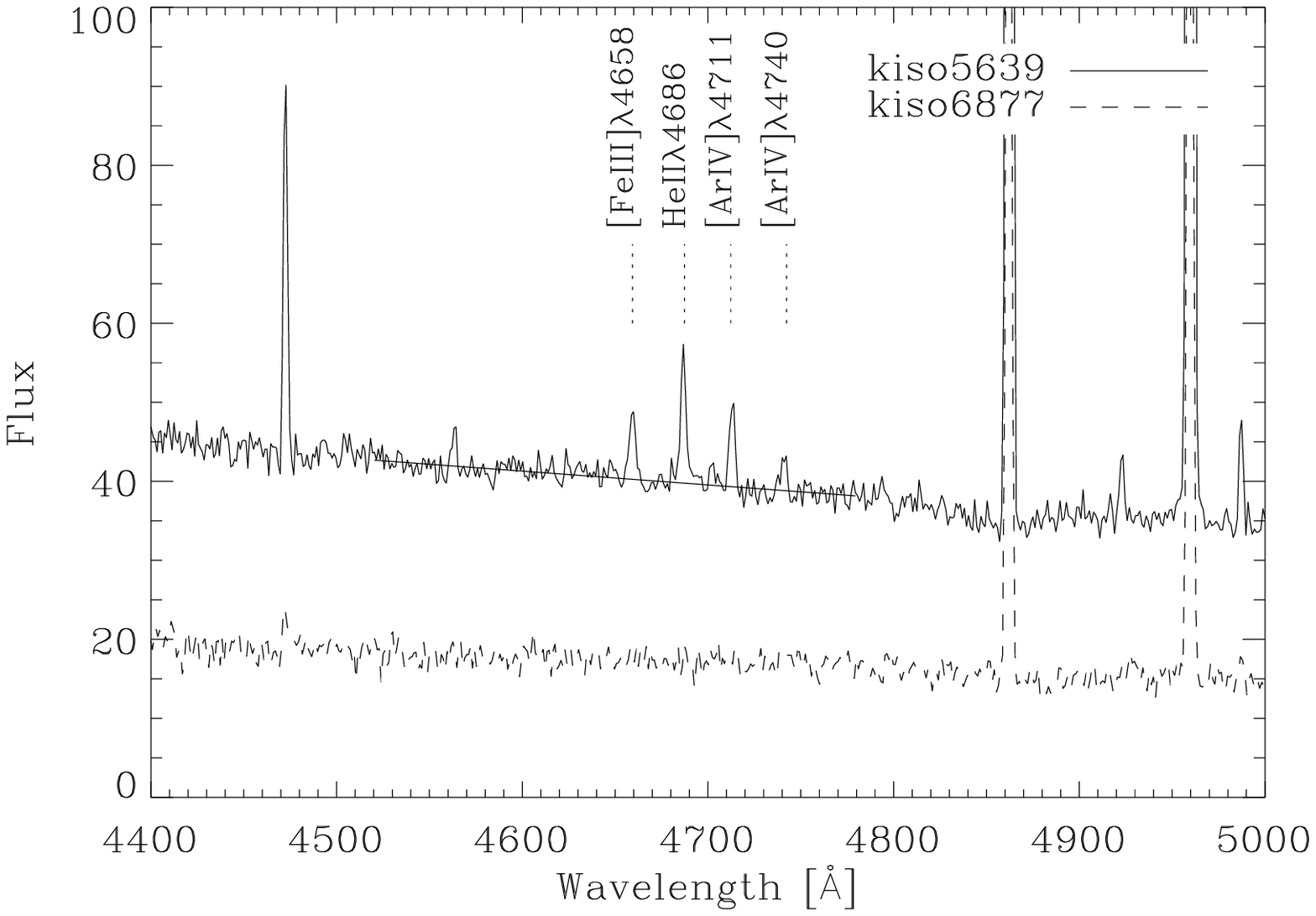}
\caption{
SDSS spectra of our two youngest targets in the wavelength
region around one of the WR bumps (4600--4680\,\AA).
The bump is missing, but one of the targets shows 
a number of high-excitation lines including HeII$\lambda$4686.
The solid slanted line represents the continuum and
is included to reinforced the lack of any bump above noise. 
}
\label{wr_features}
\end{figure}

  
Figure~\ref{equivalentw} shows the variation of
H$\alpha$ EWs across the galaxies. The largest EWs 
tend to coincide with the maximum flux 
(Fig.~\ref{brightness}), although not always  ({\sc kiso3193} 
in Fig.~\ref{equivalentw}). In general, EWs are fairly moderate,
with maxima 
$\lesssim $300\,\AA . The two exceptions correspond
to the two extremely metal poor targets to be described  
in Sect.~\ref{metalcont} -- {\sc kiso5639} and {\sc kiso6877}.
Their large H$\alpha$ EW implies the extreme youth
of the star-forming regions at the galaxy head.
We know from modeling that the 
H$\alpha$ EW of an HII region must be smaller
than some 3000\,\AA , and it drops down very quickly
so that a coeval starburst reaches EW$\,\simeq\,$200\,\AA\ in just 10\,Myr  
\citep[e.g.,][Fig.~83]{1999ApJS..123....3L}.
There is also a dependence of the EW on metallicity, but age is by far the 
dominant factor.
The large observed H$\alpha$ EWs correspond to ages of only
a few Myr, and these ages are upper limits 
to the stars responsible for the ionization since the 
(old) underlying galaxy produces continuum emission 
that reduces the observed EW. 
If {\sc kiso5639} and {\sc kiso6877} are as young as we
infer from their H$\alpha$~EWs, 
one expects to find Wolf-Rayet  (WR) star 
features in the spectra, which are characteristics
of extremely young starbursts  \citep[$<5$\,Myr; e.g.,][]{2007ARA&A..45..177C}.
These features are distinctive broad bumps at
4600--4680\,\AA\ and 5650--5800\,\AA\ 
\citep[e.g.,][]{1999A&AS..136...35S,2008A&A...485..657B}. 
Unfortunately, the WR features lie outside our
observed spectral range. However, our galaxies 
also have SDSS spectra, which cover a wider range
\citep[][]{2002AJ....123..485S,2009ApJS..182..543A}.
We inspected them for WR features but we did not see any
(see Fig.~\ref{wr_features}). Moreover, our targets were 
not found in the systematic search for WR galaxies in
SDSS carried out by \citet[][]{2008A&A...485..657B}. 
{\sc kiso6877} does not show WR features, most probably
because the SDSS spectrum was taken away from the
tadpole head, in a region which is not particularly young.
{\sc kiso5639} does not show the broad WR features either.
Instead, its spectrum contains narrow high 
excitation emission lines including HeII$\lambda$4686
(Fig.~\ref{wr_features}). These lines are supposed to be excited
only by the hard UV-radiation of WR stars, 
which makes  interpreting spectra with HeII$\lambda$4686
but without WR bumps puzzling 
\citep[see,][]{2012MNRAS.421.1043S}.
However,  there is a significant number of star-forming galaxies
without WR features showing HeII$\lambda$4686 
emission. The reason is unknown, but these galaxies 
are usually metal-poor \citep[][]{2012MNRAS.421.1043S},
so that the presence of high excitation narrow lines seems to
reflect the extreme youth of 
a metal-poor starburst. According to  \citet[][]{2012MNRAS.421.1043S},
the stellar populations at very low metallicities can have much 
higher temperatures than is currently expected in models.
Then even main sequence O stars  may excite HeII$\lambda$4686.
Alternatively, in low metallicity environments the winds
of WR stars are weak, and so,  optically thin in the He$+$ continuum, 
allowing the ionizing radiation to escape creating an HeIII region 
responsible for the observed emission
\citep[see][and references therein]{2011A&A...526A.128K}.
In short, even though {\sc kiso5639} lacks  WR bumps, 
the presence of high excitation lines such as HeII$\lambda$4686
reinforces our conclusion that its  head contains 
an extremely young metal-poor starburst.

%%%%%%%%%%%%%%%%%%
\section{Dynamical masses}\label{dynmasses}

\begin{figure}
\includegraphics[width=0.45\textwidth]{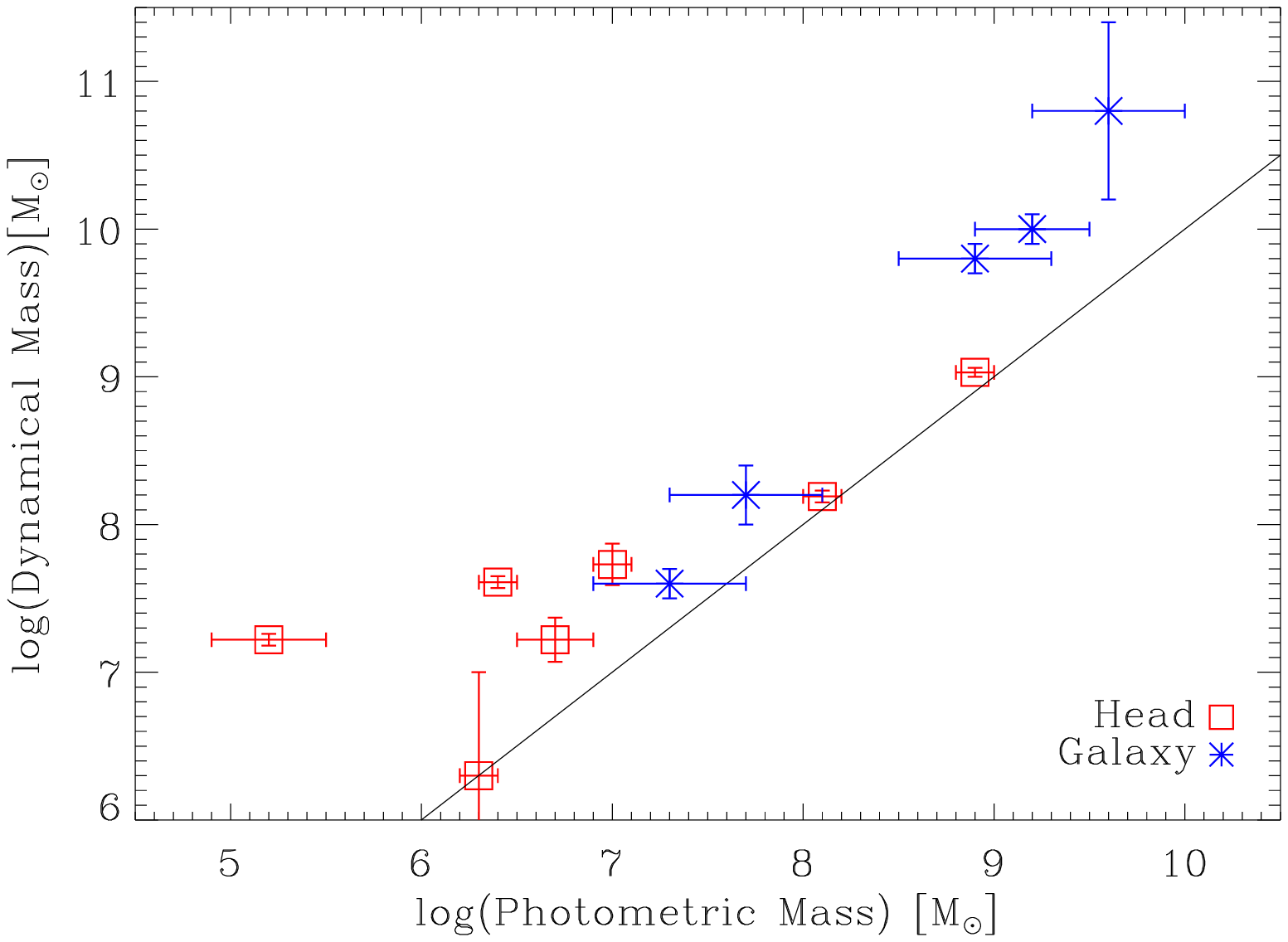}
\caption{
Dynamical mass versus photometric mass for the 
heads of the galaxies (red square symbols) and the 
full galaxies (blue asterisks).
The slanted solid line shows where the two masses are equal.
The observed dynamical masses  usually exceed 
the photometric masses. The heads of 
three tadpoles lie on the equality 
line within the errors, meaning that their 
dynamical masses are consistent with their
photometric masses.
}
\label{mass}
\end{figure}

Figure~\ref{mass} shows the relationship between
the dynamical mass and the photometric mass
for the tadpole heads and the full galaxies. The
photometric masses are from Paper~I,
whereas the dynamical masses come from applying
Eqs.~(\ref{mass_rot2}) and (\ref{veloc_turb}).
In particular, Eq.~(\ref{mass_rot2}) has been integrated  
until the last point used to compute the RC. 
%Since the galaxy extends further out, the masses thus derived
%must be regarded as lower limits.  
The actual values used for plotting are listed in Table~\ref{obs_summary}.
Note that all dynamical masses are larger than the photometric 
masses.  The difference is less important in 
the heads (except for the case of {\sc kiso3867},  
discussed in Sect.~\ref{notes_on_galax}).
%This fact just follows the general trend that the larger
%the object, the more dark matter it contains.
Three of the heads have almost identical
dynamical and  photometric masses, so in these cases
there appears to be no dark matter in the heads
({\sc kiso5149}, {\sc kiso6877}, and {\sc kiso8466}).
Keep in mind that the dynamical masses of the galaxies 
derived from RCs are actually lower limits. First, there 
is a $\sin^2i$ factor in Eq.~(\ref{mass_rot2}). 
It is probably unimportant since our galaxies are elongated
suggesting large inclinations and so  $\sin^2i\sim 1.$
Second, and more critical, is the fact that
the RC has been integrated only to the 
largest radii having velocities. 
Since the dynamical mass of the galaxies thus
derived exceeds the photometric mass, we can conclude 
that galaxies are objects with significant amounts
of non-stellar matter.

If the star-forming regions at the head of the tadpole
were self-gravitating, one would expect them to hold some 
degree of internal rotation. 
If this rotation significantly differs from the galaxy
rotation pattern, and if the head is   
massive enough, then the rotation of the head could 
perturb the rotation curve of the galaxy
producing  a noticeable  distortion
\citep[e.g.,][]{2004ApJ...611...20I}. 
We examined the observed RCs for such signals
and did not find them, except perhaps in the case of 
{\sc kiso6877}, discussed in Sect.~\ref{notes_on_galax}.

%
%%%%%%%%%%%%
%
\section{Metallicities}\label{metalcont}

\begin{figure*}
\includegraphics[width=\textwidth]{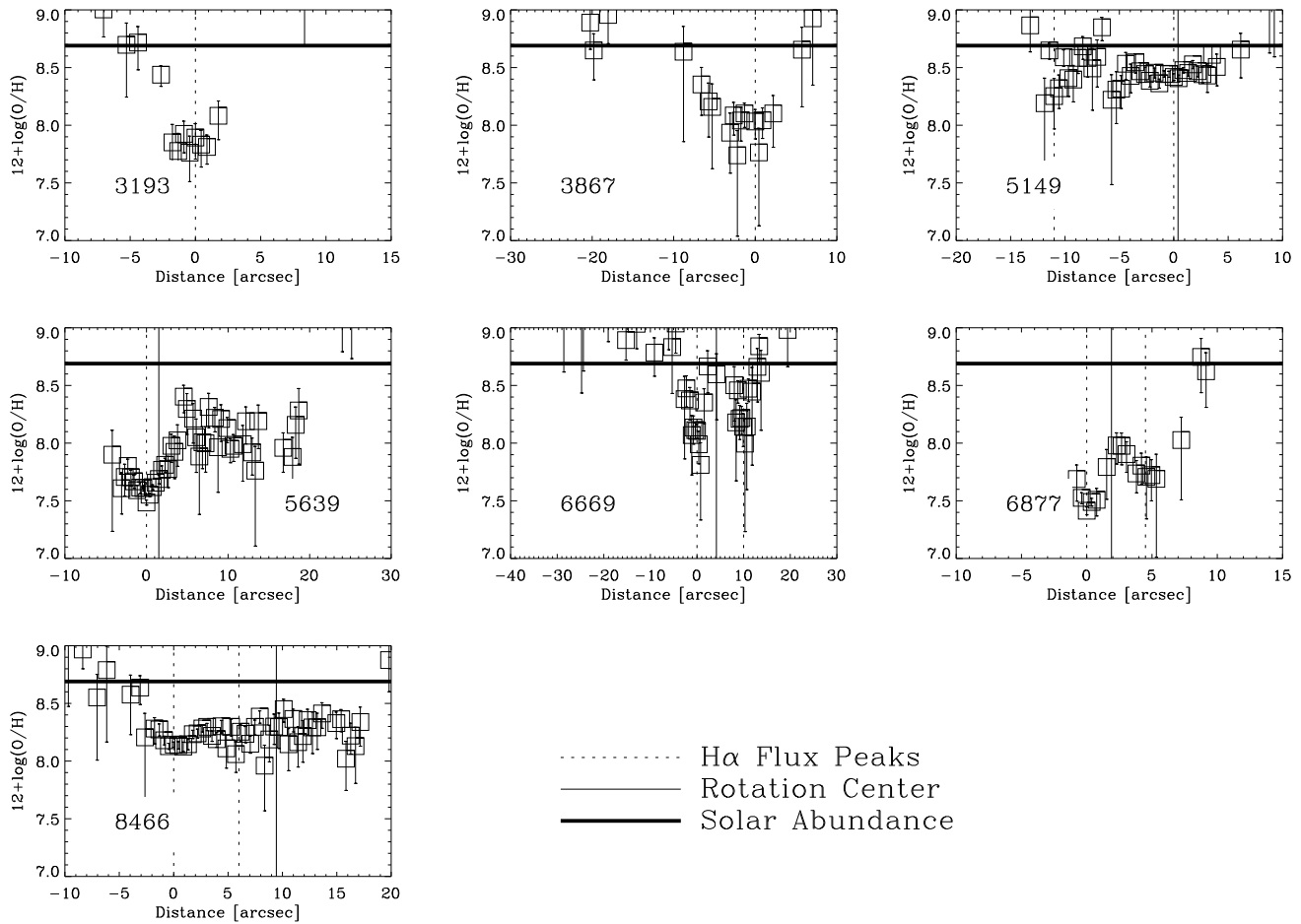}
\caption{
Oxygen abundance variation across the galaxies.
The vertical solid line represents the center
of rotation, whereas the vertical dotted lines
indicate the location of maxima in the H$\alpha$
flux profile.   
The thick horizontal solid line indicates the solar 
metallicity \citep[from][]{2009ARA&A..47..481A}.
Note the existence of abundance 
variations,
with the  minima coinciding with the largest H$\alpha$ 
signals.
Note also that {\sc kiso5639} and {\sc kiso6877} 
reach very low abundances, below one-tenth of the solar value,
therefore, they are members of the set of rare  
XMP galaxies.
The sorting of the galaxies and the  abscissae are identical 
to those in Fig.~\ref{velocity_curves}.
}
\label{metallicity}
\end{figure*}

The metallicity was estimated as explained in Sect.~\ref{determination},
using the ratio [NII]$\lambda$6583 to H$\alpha$.
Figure~\ref{metallicity} shows the variation across the galaxies 
of the oxygen abundance, including their error bars.
(Points with errors larger than 1~dex have being excluded.)
The galaxies tend to have sub-solar metallicity 
\citep[the thick horizontal line marks the solar 
oxygen abundance given by]
[$12+\log({\rm O/H})_\odot=8.69\pm 0.05$]
{2009ARA&A..47..481A}.
The galaxies also present significant abundance gradients,
with the lowest abundances tending to coincide
with the largest H$\alpha$ emissions 
(e.g., {\sc kiso6669} and {\sc kiso6877}
in Fig.~\ref{metallicity}, keeping in mind that the vertical 
dotted lines mark the position of the peak H$\alpha$ fluxes).
We also note that two targets,
{\sc kiso5639} and {\sc kiso6877},
have metallicities well below one-tenth the 
solar value, therefore, they belong to the selected 
club of XMP galaxies 
\citep[e.g.,][]{2000A&ARv..10....1K,2003A&A...407..105G}. 
They are really rare objects: one out of a thousand 
galaxies in the local universe according to    
\citet{2011ApJ...743...77M}. Therefore
the fact that we observe two in a sample of seven
cannot be a coincidence. It is  known that a significant 
fraction of XMP galaxies turn out to be cometary or tadpole
\citep{2008A&A...491..113P,2011ApJ...743...77M}.
Here we find that the reverse holds too,
i.e., that tadpole galaxies have a significant 
probability of being XMP. As we discuss in Sect.~\ref{discussion},
this fact supports the idea that the tadpole morphology is a 
sign of dynamical youth, as the 
low metallicity is a sign of being chemically young. 

\begin{figure*}
\includegraphics[width=\textwidth]{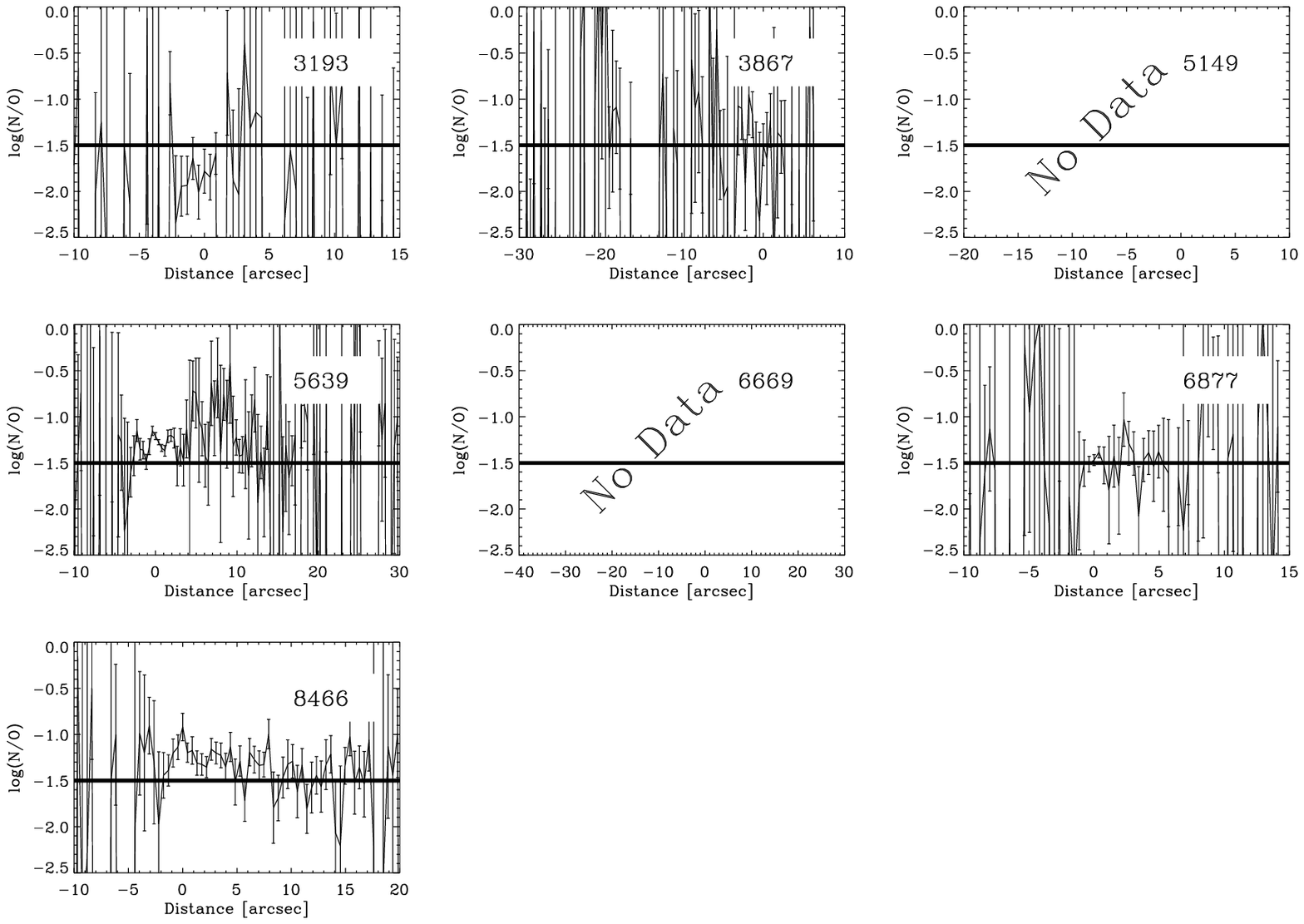}
\caption{
Variation of the ratio N to O
across the galaxies.
The horizontal solid line represents the 
plateau at some $\log({\rm N/O})\simeq -1.5$
found in most low metallicity galaxies, for which N is produced
as a primary element.
The sorting of the galaxies and the  abscissae are identical 
to those in Fig.~\ref{velocity_curves}.
The ratio could not be estimated in the two
galaxies of largest redshift, where the
required emission lines 
lie outside the observed spectral range.  
}
\label{oxovern}
\end{figure*}

The observed gradients in metallicity are one of the 
central results of this work and, therefore, 
deserve a separate discussion.  
Our abundance 
determinations are based on  N2=[NII]$\lambda$6583/H$\alpha$ 
rather than on the direct method, and this may be 
a source of systematic error \citep[e.g.,][]{2005A&A...437..849S}.
The spatial gradients in metallicity may be artificially due to
gradients in excitation. As Morales-Luis~et~al.~(2013, in preparation) 
discuss, the excitation and (to a lesser extent) N/O
change N2 at $12+\log({\rm O/H})\simeq7.5$.
The higher the excitation the smaller N2, and
the excitation is expected to change with time since the 
number of ionizing photons drops down quickly
in young  starbursts 
\citep[e.g.,][Fig.~77]{1999ApJS..123....3L}.
However, the bias the excitation produces 
is much too small to account for the $>\,$0.5~dex
gradients we detect (Fig.~\ref{metallicity}).
\citet[][Appendix~A]{2009ApJ...698.1497S}
studied the difference between the oxygen 
abundance derived from the direct method
and from N2 in a large set starburst galaxies with spectra
similar to our tadpoles.  There were no systematic 
differences within 0.2~dex for $12+\log({\rm O/H})\ge 7.7$, which
secures the reliability of the abundances found in most locations.  
As for the points with $12+\log({\rm O/H})\simeq 7.5$, N2
overestimates the oxygen abundance, which again secures the 
low metallicity values we find.
Since N2 provides the O metallicity based on the flux 
of a N line, the metallicity may be biased in 
objects with unusual N/O.
The effect of varying N/O seems to be unimportant too.
Figure~\ref{oxovern}, shows the ratio as derived
from [SII]$\lambda$6717,6731/[NII]$\lambda$6583 --
Eq.~(\ref{noequation}). (The [SII] lines are not available 
in two targets and therefore we cannot estimate their N content.)
Even though the error bars are large, the observed N/O 
looks %seems to be 
rather constant along the galaxies, at approximately the 
plateau typical of low metallicity galaxies  
\citep[$\log({\rm N/O})\simeq -1.5$; e.g., ][]{2009MNRAS.398..949P}.
Obviously, if N/O is constant then it cannot fake the observed 
metallicity  drop. However, since the error bars of N/O 
are large, we decided to run  a chi-squared test  
\citep[e.g.,][]{1986nras.book.....P} 
to further  discard  N/O as the source of the measured O
variations in {\sc kiso5639} and {\sc kiso6877}. 
We compared the observed O with the fake variations 
to be expected if O is constant but N2  varies as N/O in 
Fig.~\ref{oxovern}. The test shows how the two variations
are inconsistent with 90\% confidence.

%%%%%%%%%%%%%%%%%

%
%%%%%%%%%%%%%%%%%
%

\begin{figure}
\includegraphics[width=0.45\textwidth]{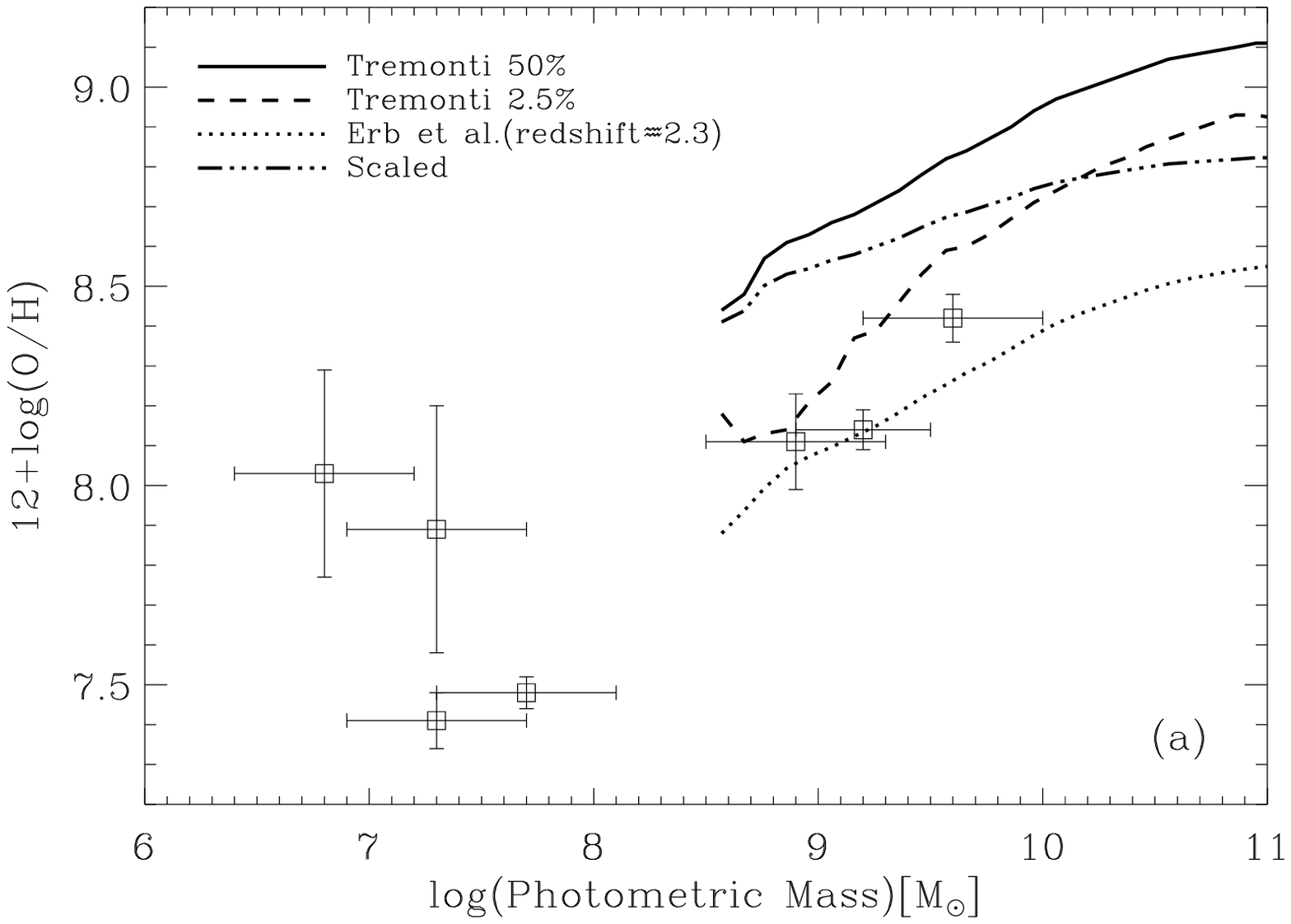}
\includegraphics[width=0.45\textwidth]{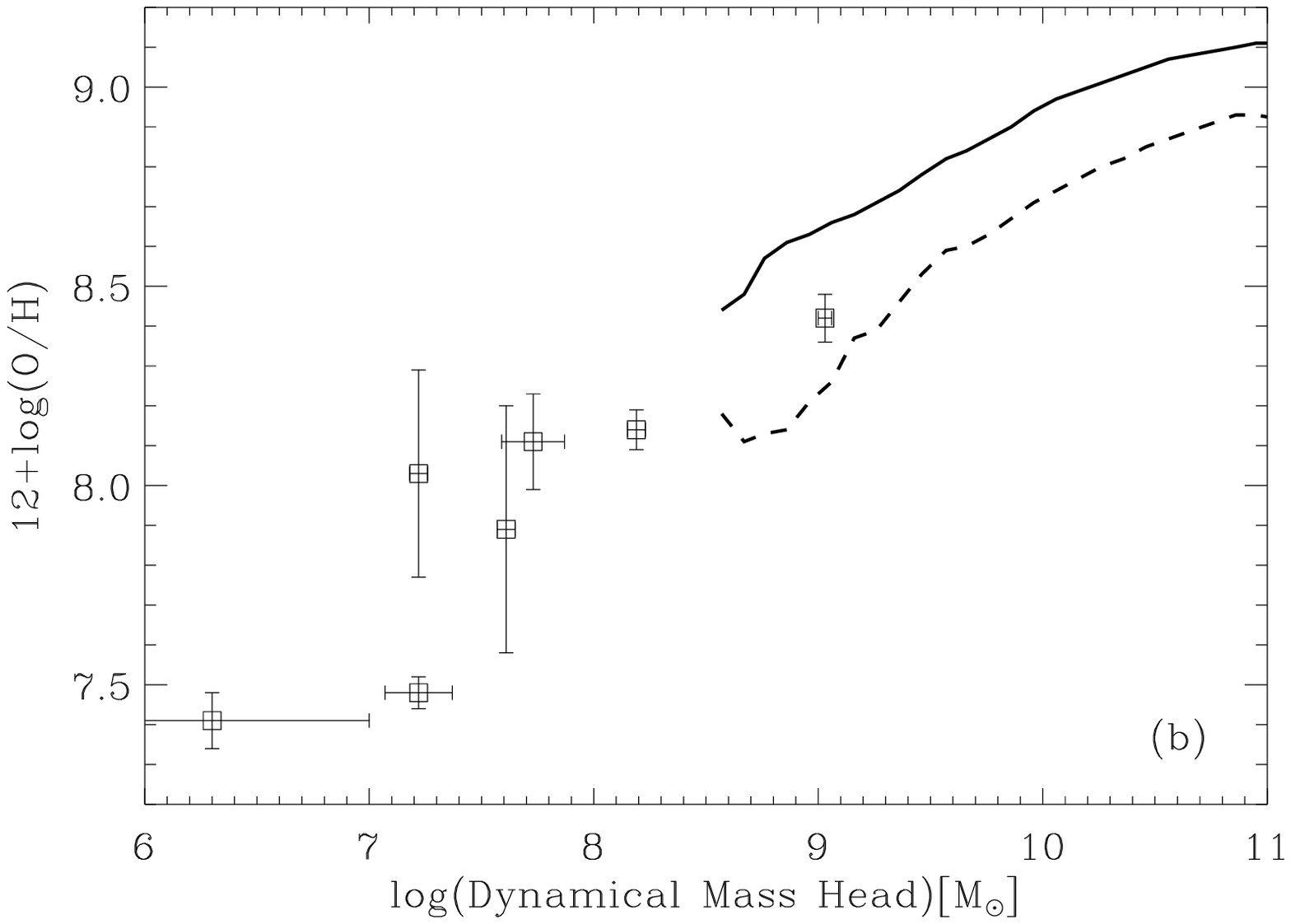}
\caption{
(a) Mass metallicity-metallicity relationship for the galaxies. 
We consider the photometric mass of the galaxy and 
the oxygen abundance at the galaxy head. 
The mean relationship found  in the local universe 
by \citet[][]{2004ApJ...613..898T} is shown as the solid and
the dashed lines -- the thick solid line represents the mean 
of the distribution, whereas the dashed line
indicates the line above which 97.5\,\% of the 
local galaxies are found. Our targets are outliers
of the local relationship; too metal poor for their masses. 
They are more in the line of the starburst galaxies found a high
redshift \citep[the dotted line, from ][]{2006ApJ...644..813E}.
The triple-dotted dashed line corresponds to the solid
line transformed to our metallicity scale 
using the prescription by \citet{2008ApJ...681.1183K}. 
(b) Same as (a) using the dynamical mass of the heads. 
The solid and dashed lines are also the same, and have 
been included for reference.
}
\label{masmetal}
\end{figure}

Figure~\ref{masmetal}a shows the mass-metallicity relationship 
for our galaxies, plotting the oxygen abundance of the 
head versus the photometric mass of the galaxy. 
The most massive tadpoles 
tend to have the largest metallicities,
although the metallicity values are 
displaced downward with respect
to the mass-metallicity 
relationship found in the local universe.
The dashed line shows the divide above which 97.5\,\% of the
local galaxies are found according to  \citet[][]{2004ApJ...613..898T}.
The tadpoles appear just below this line, so they are metal 
poor 
%in the sense argued by \citet{2009ApJ...695..259P}, i.e., 
%they are too metal poor 
for their masses\footnote{As we already 
pointed out,  the metallicity scale is not free from uncertainties.
However, even taking them into account, the tadpoles are metal 
poor. The triple-dotted dashed line in Fig.~\ref{masmetal}a
shows the solid line transformed to our metallicity
scale using the prescription by \citet{2008ApJ...681.1183K}.
The observed metallicities are well below this line too.
}.
The mass-metallicity relationship varies with redshift so that the 
same galaxy mass corresponds to lower oxygen metallicities 
at higher redshifts
\citep[a  $0.56$\,dex drop at redshift 2.3, according to][]{2006ApJ...644..813E}.
Our tadpoles are on the  mass-metallicity relationship
observed at higher redshifts (see the dotted line in Fig.~\ref{masmetal}a).
%The shift of the mass-metallicity relationship is usually
%associated with an increase of star formation rate (SFR), 
%since it is known empirically 
%that, for given mass, the larger the SFR the smaller the oxygen abundance  
%\citep{2008ApJ...672L.107E,
%2010MNRAS.408.2115M, 2010A&A...521L..53L}. 
%The explanation fits in well the type of galaxies that we analyze.
%They were selected from the Kiso catalogue of UV bright galaxies
%\citep{2010PNAOJ..13....9M}, and the UV excess is usually
%attributed to enhanced star formation activity
%\citep[e.g.,][]{2012ApJ...756..163S}. 
%
%
%%%%%%%%%%%%%%
%
%Note that we consider the metallicity at the tadpole head. This
%may look like an inconsistency but it is not so much since 
%the metallicities used to derive the local curves have been
%taking from SDSS spectra and, so, from averages
%of the central parts of the galaxies rather than the full galaxies
%\citep[][]{2004ApJ...613..898T}

We note that the metallicity of the heads also scales 
with their dynamical masses (Fig.~\ref{masmetal}b).
The relationship is even tighter than the
relationship with galaxy 
photometric
mass (Fig.~\ref{masmetal}a),
but we do not have enough points to judge whether or not 
the  improvement is statistically significant.
%If real, it may indicate that the metallicity is
%a property associated with the head, rather than
%being characteristic of the global properties of the  galaxy. 
From a practical point of view, the
reduction of scatter associated with the use 
of dynamical masses supports the reliability our 
estimate of this physical parameter.

\section{Notes on individual galaxies}\label{notes_on_galax}

The general properties of the sample are discussed in detail
in the preceding  sections. Here we focus on a few specifics
of the  individual galaxies.

\medskip
\noindent{\sc kiso3193}.
This is the only tadpole that shows no rotation, and its
line-width versus position curve has a curious double-hump 
shape (Fig.~\ref{linewidths}).  Its light distribution is less 
lopsided than for the other tadpoles 
(Figs.~\ref{images} and \ref{brightness}).
The absence of rotation combined with the increase
of linewidth toward the outskirts and the mild lopsidedness
may be consistent with a face-on disk,
and in this sense {\sc kiso3193} differs from the rest of the sample.
The observed linewidths 
are fairly large, comparable to those of  other 
larger galaxies in the sample. Consequently,
the dynamical mass of the head inferred
from linewidths is significantly larger than its
stellar (photometric) mass.

\medskip
\noindent{\sc kiso3867}. 
In both mass and size, this is the smallest tadpole 
in the sample 
(see the photometric masses of the full galaxies in
Table~\ref{obs_summary}, and also the 1kpc scales
given in Fig.~\ref{brightness}).
Its linewidths do not differ so much from the  linewidths  of other tadpoles,  
which contrasts with the low photometric mass inferred  in Paper~I
for the head of the tadpole. Consequently,
the ratio of dynamical mass to photometric mass 
is particularly high in this case -- 
%As for {\sc kiso3193}, the linewidths are as large as for the other more 
%massive tadpoles; therefore,  
the dynamical mass of the  head turns out to be  a
hundred times larger than the stellar mass
derived from photometry
(Table~\ref{obs_summary}). 
The comparatively large 
oxygen abundance we derive for the head ($12+\log({\rm O/H})\simeq 8$)
is probably an overestimate due to  observational errors. 
A few pixels away from the head the abundance drops by  0.3\,dex 
(Fig.~\ref{metallicity}), which would bring the galaxy down to
a more natural location in the mass-metallicity 
relationship (Fig.~\ref{masmetal}a).

\medskip
\noindent{\sc kiso5149}.  
The asymmetry of its RC is remarkable: see Fig.~\ref{velocity_curves}.
It extends much further out to one side,
which would imply that half of the disk is missing from
observation.   
%A large reddening  obscuring half of the disk  
%seems to be discarded  from the SDSS images,  
%which do not show a large color 
%gradient across the  galaxy.
We cannot explain how this happens,
unless the galaxy has a cigar-like shape rather
than an axi-symmetric structure.
It is unlikely that large reddening obscures half the disk, 
since the SDSS image does not show a large color 
gradient across the  galaxy.
It may be a merger, but then it would have to produce
an unlikely large scale velocity  field resembling a RC. 
The oxygen abundance is fairly constant along the 
galaxy, which seems to be associated with being a
massive object (it has the largest mass of the sample
-- Table~\ref{obs_summary}).
The galaxy presents a second bright knot separated from the
rotation center. 

\medskip
\noindent{\sc kiso5639}.
It is one of the  XMP galaxies in the sample, with the head being 
a young starburst  (a few Myr old). It also presents a very
irregular RC, particularly in the outer parts. 

\medskip
\noindent{\sc kiso6669}.
This galaxy has a second bright knot, not far from the head 
(Fig.~\ref{brightness}). The two knots  are of low metallicity, which 
contrasts with the metallicity of the rest of the 
galaxy (Fig.~\ref{metallicity}).
The center of the RC stays in between the two emission
peaks of the galaxy (and so, off-centered; 
see Fig.~\ref{velocity_curves}).
The linewidth curve shows a significant 
dip which does not coincide with any of the two
emission peaks  (Fig.~\ref{linewidths}).

\begin{figure}
\includegraphics[width=0.45\textwidth]{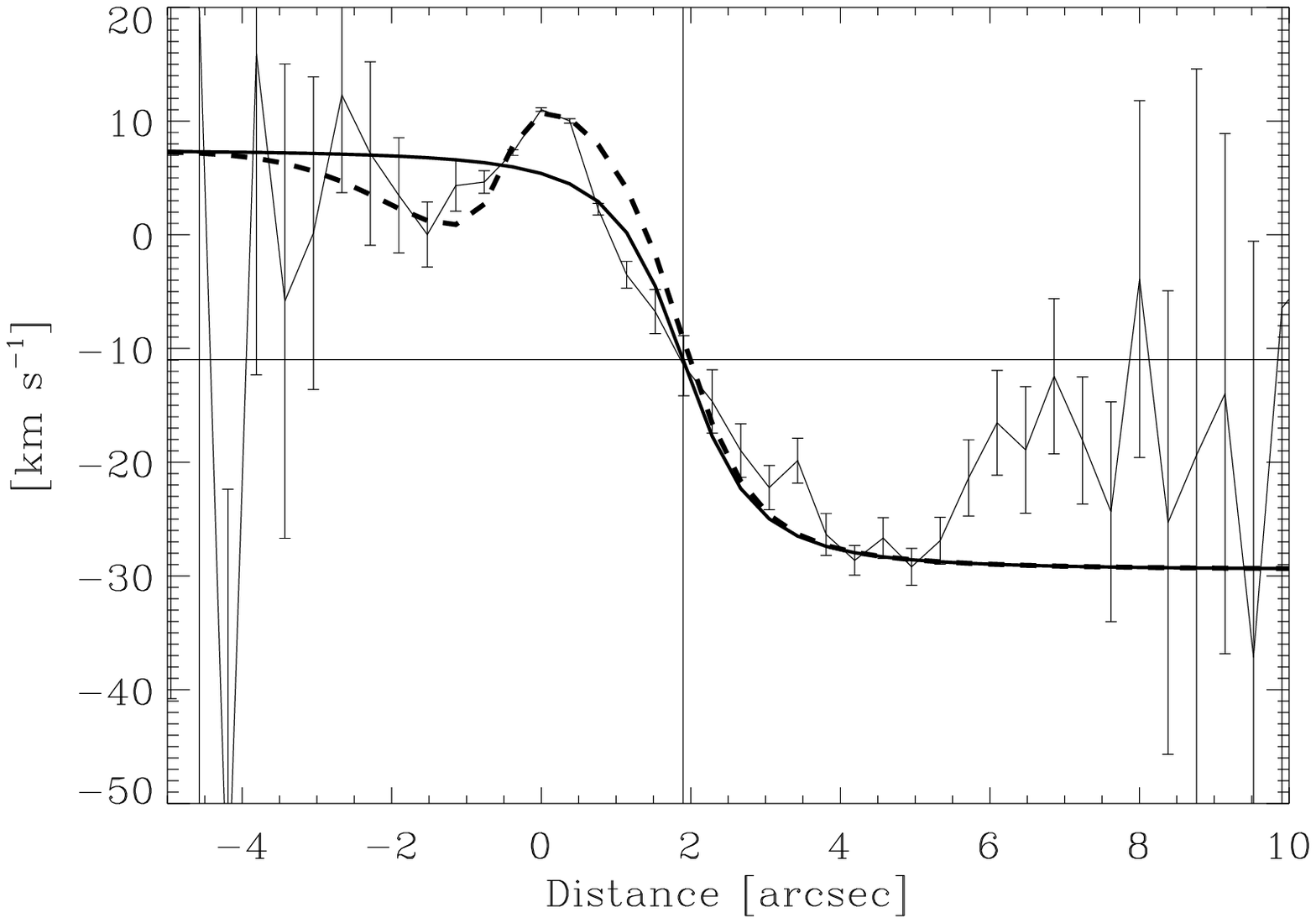}
\caption{
Velocity curve of {\sc kiso6877}  fitted by a single RC
(solid line) and  including a counter-rotating disk at the 
position of the  head (dashed line). The data points and the 
solid line are identical to those in Fig.~\ref{velocity_curves}. 
}
\label{kiso6877_spc}
\end{figure}

\medskip
\noindent{\sc kiso6877}. 
This tadpole has the lowest metallicity and is the
youngest object in the sample, and its bright head does not
seem to contain dark matter. It also presents unusual 
high-excitation emission lines in the spectral region of the 
WR bumps  (Sect.~\ref{fluxes_and_else}).
As we explain in Sect.~\ref{dynmasses}, we  sought
distortions in the RCs that could be associated with massive 
self-gravitating heads.  Perhaps {\sc kiso6877} represents the 
only example. Although we cannot assess the reliability
of the distortion present in its RC at the head position
(see Fig.~\ref{velocity_curves}), we carried out the academic exercise
of reproducing the observed velocity curve with two RCs
combined,
one for the galaxy plus one for the head. The center
of rotation of the head has been forced to be given
by the position of the head. The combined RC  
is shown as the dashed line in Fig.~\ref{kiso6877_spc}, 
and it improves the fit to the data points (compare with
the solid line).  The head is modeled with a 
counter-rotating disk with
a maximum velocity of 6\,km\,s$^{-1}$, which according to Eq.~(\ref{mass_rot2})
corresponds to 
a dynamical mass of $\log(M/M_\odot)\simeq 6.7$.
The mass thus derived is similar to both the photometric mass of the 
head and the dynamical mass inferred from linewidth
(see Table~\ref{obs_summary}).   
We also note the coincidence of this wiggle in the RC 
with an obvious decrease in line-width 
associated with the head (see Fig.~\ref{linewidths}).

\medskip
\noindent{\sc kiso8466}.
The galaxy has a multi hump structure (Fig.~\ref{images} and \ref{brightness}). 
The line-width curve presents a curious two-hump shape, which 
is not centered in photometric maxima (Fig.~\ref{linewidths}).
The oxygen metallicity is rather uniform along the galaxy, as for the other massive galaxy in the sample, {\sc kiso5149}.

\section{Discussion and conclusions}\label{discussion}

Galaxies with a bright peripheral clump on a fainter tail are called
cometary or tadpole (Fig.~\ref{images}). The origin of the shape is unknown, but
it may trace a transit phase in the assembly of many disk 
galaxies (Sect.~\ref{introduction}).
Low mass local tadpoles were identified and studied photometrically 
in Paper~I. Here we follow up the study, and analyze the chemical and 
dynamical properties of seven such targets inferred from long-slit spectra 
around H$\alpha$.

Five out the seven observed tadpoles show evidence for rotation ($\sim 70$\,\%), and  
a sixth target hints  at it. Often the center of rotation is spatially offset with 
respect  to the tadpole head (three out of five cases).
The RCs of the smaller targets are not smooth but present 
fluctuations, suggesting a complex dynamics 
(e.g., a counter-rotating  head -- see Fig.~\ref{kiso6877_spc}). 
The size and velocity dispersion of the heads are typical of  
giant HII regions and follow the scaling 
relationship known to exist between these two quantities 
(Fig.~\ref{sigmare}).
We find changes of velocity dispersion along the galaxies, 
but they are not correlated with intensity variations.
If anything, there is a tendency for the tadpole heads to 
coincide with local minima of velocity dispersion.
The head is defined as the position on the galaxy with the
largest surface brightness. Observationally, we find it 
to coincide with the region of largest H$\alpha$ flux, and
so, of largest SFR in the 
galaxy \citep[e.g.,][]{1998ARA&A..36..189K}. 
Moreover, we also find the continuum flux to extend further out
as compared to the flux in H$\alpha$, which is concentrated around the head.
Thus the bright heads seem to be  large starbursts
with their random motions reduced with respect to the 
rest of the galaxy.

Using the observed RCs and velocity dispersions, we estimate
the dynamical masses of the galaxies and their heads.
The dynamical masses of the full galaxies are between three 
and ten 
times larger than the stellar mass inferred from photometry.
The dynamical masses of the heads also exceed their
stellar masses, but to a lesser extent than the full galaxies. 
Actually, the photometric mass and the dynamical mass
of three heads 
agree within  error bars. In two other cases, however, 
the dynamical mass of the head exceeds the photometric
mass of the full galaxy.  

The  oxygen metallicity estimated from [NII]6583/H$\alpha$ often
shows significant spatial gradients  across the galaxies
($\sim$0.5~dex), being lowest at the head and increasing 
in  the rest of the galaxy, tail included. 
So far as we are aware of, this is the first time 
that a  metallicity growing away from  HII regions 
has been reported  in local galaxies. The sense of 
the resulting  metallicity gradient
is at variance with the observation of local disk galaxies, where
the gas-phase metallicity increases toward the galaxy centers
\citep{1988MNRAS.235..633V,1997ApJ...489...63G}
or is just constant \citep[][]{2012ApJ...745...66M}. 
However, the type of variation we measure,
with a minimum metallicity at the most intense star-forming region,
 has been observed in galaxies at redshift around 3 by 
\citet{2010Natur.467..811C} where it is interpreted as evidence 
for infall of  pristine gas triggering star formation. 
Once systematic errors are disregarded
(as we did in Sect.~\ref{metalcont}), 
it is difficult to avoid such interpretation, also in the case of our 
tadpole galaxies. 
We considered and then discarded the following two 
possibilities: (1) assume a regular metallicity distribution decreasing
outward. If the head was formed from gas in the 
galaxy outskirts, but has spiraled in toward the 
center by dynamical friction \citep[e.g.,][]{2012ApJ...747..105E},
then the head would naturally present a metallicity lower than its 
immediate surroundings and similar to the galaxy outskirts. 
However, this prediction of metallicity gradients
induced by internal migration is inconsistent with 
our observation, where the metallicity 
at the head is not just lower than the surroundings 
but the lowest (see Fig.~\ref{metallicity}).
(2) Metal-rich supernova (SN) driven winds remove metals
from shallow gravitational potential wells,
producing metal-poor galaxies  
\citep[e.g.,][]{1999ApJ...513..142M,2004A&A...426...37R}. 
This mechanism explains the overall low gas metallicity of some 
galaxies with significant old stellar populations, but it does not
account for the presence of a region like the tadpole head, with 
metallicity lower than the rest of the galactic gas.  
The metals ejected by the winds are those 
created by the stars that explode as SNe. They are not 
the metals of the ambient gas. Then the loss of these ejecta 
reduces the efficiency of the galaxy to retain metals, 
but it does not reduce the metallicity of a particular
region of the galaxy.
In short,  the interstellar medium of the
tadpole  galaxies is  not well mixed but shows significant metallicity gradients.
Since the mixing time scale is expected 
to be relatively short  \citep[shorter than a few Myr; e.g.,][]{1996AJ....111.1641T,2002ApJ...581.1047D},  
the recent infall of metal-poor gas seems to be the only viable 
alternative to explain the metallicity drop observed at the 
tadpole heads. 

The observationally motivated interpretation
of external gas infall fits in well the 
cold-flow gas accretion scenario arising from 
cosmological numerical simulations 
\citep[e.g.,][]{2005MNRAS.363....2K,2009Natur.457..451D}.
It predicts localized accretion in clumpy streams of pristine
gas ready to create stars. The streams may directly form giant clumps 
that we detect as tadpole heads or,
alternatively, feed the disks with turbulent gas that 
eventually fragments into giant clumps by gravitational instability.
In both cases the massive clumps are prone to migrate toward the
galaxy centers and become progenitors of central
spheroids \citep{2008ApJ...688...67E,2010MNRAS.404.2151C}.
Other details of the observed tadpole properties are also consistent
with the cold-flow accretion scenario. The process is expected to be
ubiquitous at high redshift. Then the flows fade away gradually in 
a process that has not being completed in small galaxies yet 
\citep[e.g.,][]{2009MNRAS.395..160K}.
This prediction is consistent with the absence of low metallicity
HII regions in large local spirals, as well as in our most
massive tadpoles ({\sc kiso5149} and {\sc kiso8466}; see 
Fig.~\ref{metallicity}), for which the infall would be already over. 
The metallicity drop is observed only in the low mass objects, 
reflecting the downsizing process in galaxy formation.

The geometrical displacement 
of some of the tadpole heads with respect to the
centers of rotation also favors the cold-flow scenario. 
The expected streams of cold gas never end up at the galaxy 
center. The clumps are formed in the disk, and require 
time to be transported inward.

Extremely metal poor (XMP) galaxies are rare. Tadpole
galaxies are also rare. The fact that we observe two XMP
galaxies in a sample of seven tadpoles cannot be  casual
(Sect.~\ref{metallicity}).  It is known that a significant fraction 
of XMP galaxies turns 
out to be tadpole or cometary \citep{2008A&A...491..113P,2011ApJ...743...77M}.
Here we find the reverse to be true as well, i.e., tadpoles have 
a significant chance of being XMP. The coincidence of these
two seemingly disconnected properties is best understood 
if the objects are primitive, with the cometary shape and
the low metallicity reflexing dynamical and chemical youth, 
respectively.

%The observed tadpoles follow a mass-metallicity relationship,
%with the more massive ones having larger oxygen abundance.
%However, they are low metallicity outliers of the local 
%mass-metallicity  relationship. We interpret the 
%low metallicity for a given mass as an effect the large 
%specific SFR of the tadpoles -- it is known that
%the larger the SFR the smaller the metallicity
%\citep{2008ApJ...672L.107E,
%2010MNRAS.408.2115M, 2010A&A...521L..53L}. 

All these results combined are consistent with the local 
tadpole galaxies being  turbulent disks in early stages of assembling. 
Their star formation seems to be sustained by  
accretion of external metal-poor gas.

\acknowledgements
%
% Debra and Bruce said explicitly that they do not have  acknowledgements 
% for this paper.
%
Thanks are due to 
R.~Amor\'\i n, 
J.~Rodr\'\i guez Zaur\'\i n,
R.~S\'anchez-Janssen,
G.~Stinson,
R.~Terlevich, 
J.~V\'\i lchez 
and E.~Wisnioski
for  enlightening discussions and 
suggestions on various aspects  of the work.
Thanks are also due to the anonymous referee for comments
that help improving the presentation and the discussions 
in the paper.
A.~Varela kindly provided the values for the seeing during 
observation included in Table~\ref{logbook}.
This work has been partly funded by the Spanish Ministry for Science, 
project AYA~2010-21887-C04-04.          % new
%
% Consolider GTC
JSA, CMT and JMA are members of the Consolider-Ingenio 2010 Program, grant 
MICINN CSD2006-00070: First Science with GTC.
The article is partly based on observations made with the telescopes INT and NOT 
operated at the Spanish {\em Observatorio del Roque de los Muchachos} (ORM).
%Add ack to the ORM telescopes, and also to the use of service time of the 
%NOT (both NOT service and spanish service ...).
%
%The IAC support  astronomer crew (Pilar Monta\~nez), that observed.
%
{\sc kiso5639} and {\sc kiso6669} were observed through the 
NOT Fast-Track Service Program, whereas {\sc kiso6877} corresponds
to Spain Service Time at ORM.
The instrument ALFOSC is provided by the Instituto de Astrof\'\i sica de 
Andaluc\'\i a (IAA) under a joint agreement with the University of 
Copenhagen and NOTSA.
Funding for SDSS, SDSS-II, and  SDSS-III has been provided by the Alfred P. Sloan Foundation, 
the Participating Institutions, the National Science Foundation, 
and the U.S. Department of Energy Office of Science.
%
%%%%%%%%%%%%%%%
%
\newcommand\nar{New Astron. Rev.}
%\bibliography{ms,/Users/jos/texto/papers/references/galax}
%\bibliographystyle{aa}%{apj}

\end{document}